\documentclass[trackchanges,twocolumn]{aastex701}
\usepackage{natbib}
\usepackage{color}
\usepackage{mathtools}
\usepackage{booktabs}
\usepackage{hyperref} 

\def    \apjl  		{\rm {ApJL}}

\def    \apjl  		{\rm {ApJL}}

\def	\cm		{\,{\rm {cm}}}
\def	\K		{\,{\rm K}}
\def	\g		{\,{\rm {g}}}
\def	\mum	{\,{\mu \rm{m}}}

\def \bea {\begin{eqnarray}}
\def \ena {\end{eqnarray}}                  

\def	\ba	{\boldsymbol{a}}

\def 	\bE	{\boldsymbol{E}}

\def	\C	{{\rm C}}

\def	\cm	{\,{\rm cm}}

\def	\d	{{\rm d}}

\def	\D	{{\rm D}}

\def	\g	{\,{\rm g}}
\def	\gas	{\,{\rm gas}}
\def	\G	{{\rm G}}

\def	\H	{{\rm H}}

\def	\N	{{\rm N}}

\def	\s	{\,{\rm s}}

\def	\V	{{\rm V}}

\def    \bE     	{\boldsymbol{E}}

\newcommand{\alfvenic}{Alfv$\acute{\text{e}}$nic~} 

\begin{document}
\shorttitle{3D B-fields in G11.11-0.12}
\shortauthors{Truong et al.}
\title{3D B-fieLds in the InterStellar medium and Star-forming regions (3D-BLISS): I. Using Starlight Polarization in the Massive IRDC Filament G11.11-0.12}

\author[0000-0001-9654-8051]{Bao Truong}
\affiliation{Korea Astronomy and Space Science Institute, Daejeon 34055, Republic of Korea} 
\email[show]{baotruong@kasi.re.kr}
\affiliation{Department of Astronomy and Space Science, University of Science and Technology, 217 Gajeong-ro, Yuseong-gu, Daejeon, 34113, Republic of Korea}

\author[0000-0003-2017-0982]{Thiem Hoang}
\affiliation{Korea Astronomy and Space Science Institute, Daejeon 34055, Republic of Korea} 
\email{thiemhoang@kasi.re.kr}
\affiliation{Department of Astronomy and Space Science, University of Science and Technology, 217 Gajeong-ro, Yuseong-gu, Daejeon, 34113, Republic of Korea}

\author[0000-0002-5913-5554]{Nguyen Bich Ngoc}
\email{capi37capi@gmail.com}
\affiliation{Vietnam National Space Center, Vietnam Academy of Science and Technology, 18 Hoang Quoc Viet, Hanoi, Vietnam}

\author[0000-0002-3681-671X]{Nguyen Chau Giang}
\affiliation{Korea Astronomy and Space Science Institute, Daejeon 34055, Republic of Korea} 
\email{chaugiang@kasi.re.kr}
\affiliation{Department of Astronomy and Space Science, University of Science and Technology, 217 Gajeong-ro, Yuseong-gu, Daejeon, 34113, Republic of Korea}

\author[0000-0002-6488-8227]{Le Ngoc Tram}
\email{lengoctramlyk31@gmail.com}
\affiliation{Leiden Observatory, Leiden University, PO Box 9513, 2300 RA Leiden, The Netherlands}

\author[0000-0003-1990-1717]{Ng\^{a}n L\^{e}}
\email{nganle@kasi.re.kr}
\affiliation{Korea Astronomy and Space Science Institute, Daejeon 34055, Republic of Korea}

\begin{abstract}
Measuring three-dimensional magnetic fields (3D B-fields) is essential to understand the formation and evolution of the interstellar medium and multi-scale star formation; however, the accurate measurement of 3D B-fields is still challenging. The dust polarization angles by magnetically aligned grains provide the projected B-fields onto the plane-of-sky, while the dust polarization degree provides the B-field's inclination angle with respect to the line-of-sight. Our previous theoretical studies proposed a new method of probing 3D B-fields using dust polarization combined with the Radiative Torque (RAT) alignment theory and demonstrated the accurate inference of B-field inclination angles using synthetic polarization data. In this paper, we report the first application of the new technique to study 3D B-fields and dust properties in the G11.11-0.12 filament (hereafter G11) from starlight polarization observations taken by ISRF/SIRPOL at $2.19\,\rm\mu m$. Using both observed starlight polarization and optical dust extinction curves from the Gaia mission, we constrained the maximum grain size of $0.25\,\rm\mu m$ and the grain elongation with an axial ratio of $s\gtrsim 1.4$ in the outer regions of G11. We calculated the alignment properties in G11 by using the \textsc{DustPOL\_py} code. The B-field's inclination angles in G11 are then inferred from the observed starlight polarization efficiency when the grain alignment is included, with a mean angle of $\sim 48$ degrees. From these inferred inclination angles, we found evidence of the local 3D arc-shaped B-field structure toward the sightline. These findings are important for understanding 3D B-field's roles in the formation and evolution of massive filamentary clouds.   
\end{abstract}



\section{Introduction}
Magnetic fields (hereafter B-fields) are an important physical parameter in shaping the structure and driving the evolution of the interstellar medium (ISM) and the star formation process. To accurately understand the dynamical roles of B-fields in the ISM evolution and star formation, their strength and morphology in a three-dimensional (3D) space must be determined (i.e., 3D B-fields,\; see the review of \citealt{Tahani2022_review}). The measurement of 3D B-fields is important to investigate multi-scale star formation in a magnetized medium: testing the large-scale filamentary cloud formation scenario by multi-shock compression (\citealt{Inutsuka2015}; \citealt{Inoue2018}; \citealt{Tahani2018, Tahani2022}), probing small-scale filament evolution by cloud-cloud collisions (\citealt{Maity2024}); or testing the hourglass-shaped B-field pattern in the case of magnetically-regulated protostellar core formation (\citealt{Kandori2017II, Kandori2018IV}; \citealt{Basu2024}). 

The strength and morphology of B-fields in the ISM and star-forming regions have been measured through Zeeman splitting of different spectral lines (e.g., HI, OH, CN and CCS, see \citealt{Crutcher1999}; \citealt{Falgarone2008}; \citealt{Crutcher2012}), the polarization of starlight by dichroic extinction at optical and near-IR (NIR) wavelengths $\lambda \sim 0.1 - 20\,\rm\mu m$ (\citealt{Hall1949}; \citealt{Hiltner1949}; \citealt{Andersson2015}) and thermal dust polarization by dichroic emission in the far-IR and sub-mm wavelengths $\lambda \sim 200 - 1000\,\rm\mu m$ (\citealt{Hildebrand1988}; \citealt{Planck2015}; \citealt{Pattle2019}), Faraday rotation (\citealt{Tahani2018, Tahani2022}), and synchrotron polarization (\citealt{Beck2015}; \citealt{Padovani2018,Padovani2021b, Padovani2021a}). These B-field techniques, however, can only trace one dimension of B-fields. For instance, the Zeeman splitting of spectral lines and Faraday rotation could only provide the 1D component of B-fields along the line-of-sight (LOS, i.e., $B_{\rm LOS}$), while the orientation of starlight and thermal dust polarization, and synchrotron polarization could only provide the 2D orientation of B-fields in the plane-of-sky (POS, i.e., $B_{\rm POS}$). There are proposed techniques to probe 3D B-fields in star-forming regions: by combining Faraday rotation and thermal dust polarization (\citealt{Tahani2018, Tahani2022}); by combining Zeeman measurements and dust polarization (\citealt{Reissl2018}; \citealt{Hwang2024}); or by the velocity gradient technique (VGT, see, e.g., \citealt{Casanova2017}; \citealt{Hu2023a,Hu2023b}) with the help of Artificial Neural Networks (\citealt{Hu2024}; \citealt{Hu2025}). 

Dust polarization from differential extinction and emission of magnetically aligned grains not only provides us with the POS B-field patterns, but their polarization degree is also a powerful tool to obtain information about the alignment and the physical properties (size, shape, and composition) of dust grains. While the usage of the polarization angle to trace B-fields has been widely used, the usage of the polarization degree to probe dust physics and dust properties is still far from complete. One of the main reasons is the effects of competing processes on the observed polarization degree. In the diffuse ISM with low gas column density of $N_{\rm H} \lesssim 1.5 \times 10^{22}\,\cm^{-2}$ where grain alignment is efficient, the observed polarization degree is dominantly influenced by the fluctuations of B-fields along the sightline by magnetic turbulence (\citealt{Planck2015, Planck2020}; \citealt{Angarita2023}). In dense molecular clouds, filaments, and starless cores with $N_{\rm H} \gtrsim 1.5 \times 10^{22}\,\cm^{-2}$, the depolarization mechanism is more complicated due to the joint action of B-field geometries and the reduction of the grain alignment efficiency (see \citealt{King2019}; \citealt{Hoang2021}; Giang et al. in preparation).

The wavelength-dependent polarization spectrum provides crucial constraints on dust composition by astrosilicate and carbonaceous components (e.g., separate or composite grain models; see \citealt{Draine2009}; \citealt{Guillet2018}; \citealt{Draine2021}), while the maximum polarization degree provides constraints on the axial ratio of the grain shape (hereafter grain elongation), porosity (see \citealt{Draine2021b, Draine2021}; \citealt{Draine2024a, Draine2024b}), and grain alignment efficiency \citep{Lee2020}. The observed starlight/thermal dust polarization degree can be a diagnosis of grain growth in dense star-forming regions. The increasing peak wavelength $\lambda_{\rm max} > 0.55\,\rm\mu m$ of starlight polarization spectrum at high-extinction regions $A_{\rm V} > 4$ indicates the evidence of grain growth with the maximum grain size $a_{\rm max} > 0.25\,\rm\mu m$ (\citealt{Whittet2008}; \citealt{Vaillancourt2020}). Meanwhile, the profiles of the starlight polarization efficiency $P_{\rm ext}/A_{\rm V} $ versus $A_{\rm V}$ (i.e., $P_{\rm ext}/A_{\rm V} \varpropto A^{-\alpha}_{\rm V}$), or the thermal dust polarization degree $P_{\rm em}(\%)$ versus $A_{\rm V}$ (i.e., $P_{\rm em}(\%) \varpropto A^{-\alpha}_{\rm V}$), could diagnose the presence of large grains in dense filaments and starless cores through the observed slope $\alpha$ or the maximum visual extinction $A_{\rm V, max}$ where grain alignment is still efficient (see \citealt{Whittet2008}; \citealt{Hoang2021}; \citealt{TruongHoang2025}; \citealt{Tram2025}). 

Recently, a significant effort has been invested to construct 3D B-fields using only dust polarization, through combining 2D dust polarization angles and the inclination angle constrained by the polarization degree. \cite{Chen2019} first used thermal dust polarization fraction $P_{\rm em}(\%)$ to infer the 3D B-field inclination angle $\gamma$ (i.e., the angle between the mean B-field and the LOS) in Vela C cloud observed by Balloon-borne Large Aperture Submillimeter Telescope for Polarimetry (BLASTPol, see \citealt{Chen2019}). \cite{HuLaz2023} improved the new technique by incorporating anisotropic magnetohydrodynamic (MHD) turbulence and showed that Chen's technique can only be valid when the cloud is sub-\alfvenic. For starlight polarization, the effect of inclined B-fields is characterized from the observations the starlight polarization degree per reddening, denoted by $P_{V}/E(B-V)$. \cite{Panopoulou2019b} conducted optical polarimetric observations toward 22 stars in the region with the highest thermal dust polarization degree $p_{\rm 353GHz} \approx 20\%$ from Planck observations (\citealt{Planck2020}), and found starlight polarization degree reddening $(P_{V}/E(B-V))$ higher than $13\%\,\rm mag^{-1}$. \cite{Angarita2023} demonstrated the upper bound of $(P_{V}/E(B-V)) \approx 16\%\,\rm mag^{-1}$ when magnetic fields are mostly lying on the POS. Note that for tracing 3D B-fields from both starlight and thermal dust polarization studies, the grain alignment is considered to be perfect within the clouds (see \citealt{Chen2019}; \citealt{HuLaz2023}; \citealt{Angarita2023}; \citealt{Doi2024}). Therefore, a detailed study of grain alignment physics is required to accurately interpret the polarization from absorption and emission of aligned grains and the application of the dust polarization degree to probe 3D B-fields and dust physics in various astrophysical environments.


The leading theory of grain alignment is based on Radiative Torques (RATs, see \citealt{Lazarian2007,Hoang2008}). The alignment by RATs was numerically tested for various environmental conditions (\citealt{Hoang2014}) and dust compositions (\citealt{Hoang2016a}), and provided quantitative predictions of the variation of starlight and thermal dust polarization in star-forming regions (see reviews of \citealt{Andersson2015, Tram2022}). The extended RATs theory by unifying with magnetic relaxation (a.k.a Magnetically Enhanced RAdiative Torque or MRAT, see \citealt{Hoang2016a}) demonstrated the importance of embedded iron clusters inside grains that enhance magnetic alignment by RATs (see also \citealt{Hoang2022}; \citealt{Giang2023}). 



Based on the modern physics of grain alignment by RATs, \cite{HoangTruong2024} have developed a new technique for probing 3D B-fields using the dust polarization degree. By including the alignment by RATs, magnetic turbulence and dust properties effects, the inclination angles can be accurately inferred from the thermal dust polarization degree, which was tested on synthetic polarization observations of filamentary clouds. \cite{TruongHoang2025} have extended the new technique to probe 3D B-fields from starlight polarization efficiency at optical/NIR wavelengths. This unlocks a new possibility of characterizing 3D B-fields and dust properties in multi-scale star-forming regions by using multi-wavelength starlight and thermal dust polarization observations.

In this study, for the first time, we apply our proposed technique to polarimetric observations of the Infrared Dark Cloud (IRDC) filament G11.11-0.12 (hereafter G11). It is a dense, cold filamentary cloud located in the Galactic plane ($l = 11.^{\circ}119$ and $b = -0.^\circ 0647$, see \citealt{Wang2014}) at a distance of 3.6 kpc (\citealt{Pillai2016}). The filament is massive, with the linear mass density of $\sim 600\,\rm M_{\odot}/pc$ (\citealt{Kainulainen2013}), and consists of massive clumps and protostellar cores formed and evolved along the filament spine (\citealt{Henning2010}; \citealt{Kainulainen2013}; \citealt{Wang2014}). Two massive clumps, P1 and P6, are considered as sites of high-mass star formation and were identified by James Clark Maxwell Telescope/Submillimetre Common-User Bolometer Array (JCMT/SCUBA) at $850\,\rm\mu m$  (\citealt{Henning2010}) with a mass of $\sim 1000\,\rm M_{\odot}$. The presence of POS B-fields is identified in the outer regions of G11 from starlight polarization observations at $2.19\,\rm\mu m$ by \cite{Chen2023} and in the inner dense filament by Stratospheric Observatory for Infrared Astronomy/High-resolution Airborne Wide-band Camera (SOFIA/HAWC+) at $214\,\rm\mu m$ (\citealt{Ngoc2023}) as parts of the FIELDMAPS legacy project (PI: Ian W. Stephens, Worcester State University). Both observations demonstrated the perpendicular orientation of POS B-fields to the filament spine and the dominant B-field roles in the early phases of massive filament formation and evolution (i.., sub-\alfvenic turbulence). This makes G11 an ideal candidate for examining the effectiveness of the new technique to probe 3D B-field and dust properties in this environment.

This is the first paper in our series aiming to investigate 3D B-fields in the Interstellar medium and Star-forming regions (3D-BLISS). We first apply the new technique to the available starlight polarization observations by \cite{Chen2023} at $2.19\,\rm\mu m$ that mostly obtain POS B-fields in the outer regions of G11 with the filament width of 5 pc (\citealt{Kainulainen2013}). We utilize the latest version of \textsc{DustPOL\_py} code that has incorporated the numerical modeling of dust extinction and polarization based on the modern physics of RAT alignment mechanism (\citealt{Lee2020}; \citealt{Tram2021, Tram2024, Tram2025})\footnote{https://github.com/lengoctram/DustPOL\_py} to characterize the major effects of grain alignment and dust properties within G11. We infer the local 3D inclination angles from the observed starlight polarization efficiency using the new technique developed by \cite{TruongHoang2025}, and investigate the applications in reconstructing 3D B-field morphology and understanding its importance in filament formation and evolution. The study of 3D B-fields in the inner dense filament from thermal dust polarization observations will be presented in our second paper of the 3D-BLISS series (Ngoc et al., in preparation).

The structure of this paper is as follows. Section \ref{sec:method} reviews the technique of probing local inclination angles of the mean B-fields from the starlight polarization developed by \cite{TruongHoang2025}. Section \ref{sec:obs} presents the archival starlight polarization observations by \cite{Chen2023}, and the characterization of the observed starlight polarization efficiency and the effect of magnetic turbulence.  The constraints on intrinsic dust properties and grain alignment, followed by the calculation of the intrinsic polarization efficiency and polarization coefficient fraction  will be in Sections \ref{sec:dust_align} and \ref{sec:int_fpol}. The results of inferred inclination angles and 3D B-field in G11 are presented in Section \ref{sec:result_incl}. Section \ref{sec:discuss} demonstrates the implications of 3D inclination angles for studying 3D B-field strength and morphology of G11 for understanding filament evolution. The conclusion of our main findings is summarized in Section \ref{sec:summary}. 

\section{Technique Overview}
\label{sec:method}

In this section, we present an overview of the new technique of inferring 3D inclination angles from starlight polarization by \cite{TruongHoang2025}. Grains with irregular shapes aligned with B-fields can attenuate radiation from background starlight at optical-NIR wavelengths. The transmitted starlight extinct by magnetically aligned grains is polarized with $\bf P \parallel B$ (see \citealt{Andersson2015}). In the regime where polarized light is optically thin (i.e., the polarized optical depth of $\tau_{\rm pol} \ll 1$), the wavelength-dependent starlight polarization degree normalized to the hydrogen column density (a.k.a. polarization efficiency), denoted by $P_{\lambda}/N_{\rm H}$, can be described as (see \citealt{TruongHoang2025})

\bea
\frac{P_{\lambda}}{N_{\rm H}} &=& \sin^2\gamma F_{\rm turb}\int_{a_{\rm min}}^{a_{\rm max}}f_{\rm align}(a)C_{\rm pol}(a,\lambda) n_{d}(a)da\nonumber\\
&=&  \sin^2\gamma \times  F_{\rm turb} \times \sigma_{\rm pol, align}(\lambda),\label{eq:Pmod}
\ena
where 
\bea
\sigma_{\rm pol, align}(\lambda) &=& \int_{a_{\rm min}}^{a_{\rm max}}f_{\rm align}(a)C_{\rm pol}(a,\lambda) n_{d}(a)da,\label{eq:sigma_pol_align}
\ena
is the polarization coefficient of aligned grains. $\gamma$ is the inclination angle of mean B-fields with respect to the sightline and $F_{\rm turb}$ is the factor describing the impact of magnetic turbulence on the starlight depolarization (see Appendix \ref{sec:appendix_Fturb}). $n_d(a)$ is the grain size distribution from the lower ($a_{\rm min}$) to the upper limit ($a_{\rm max}$) of the effective grain size $a$, and $f_{\rm align}(a)$ is the alignment function with respect to the grain size $a$ (see Equation \ref{eq:falign}). $C_{\rm pol}(a, \lambda)$ is the polarization cross-section of the oblate grains and is calculated by
\begin{equation}
    C_{\rm pol}(a, \lambda) = \frac{1}{2}[C_{\rm ext}(a, \lambda)(\bE \perp \ba_{1}) - C_{\rm ext}(a, \lambda)(\bE \parallel \ba_{1})],
\end{equation}
where $C_{\rm ext}(a, \lambda)(\bE \perp \ba_{1})$ and $C_{\rm ext}(a, \lambda)(\bE \parallel \ba_{1})$ are the extinction cross-sections for the incident electric fields $\bE$ perpendicular and parallel to the grain axis of symmetry $\ba_{1}$, respectively. In this study, the values of $C_{\rm ext}(a, \lambda)(\bE \perp \ba_{1})$ and $C_{\rm ext}(a, \lambda)(\bE \parallel \ba_{1})$ for different grain sizes, elongations, and wavelengths are directly obtained from the Astrodust model (see \citealt{Draine2021}; \citealt{Hensley2023})\footnote{http://arks.princeton.edu/ark:/88435/dsp01qb98mj541}.

The polarization efficiency can also be rewritten in short form as
\begin{equation}
    \frac{P_{\lambda}}{N_{\rm H}} = \frac{P_{\rm i}(\lambda)}{N_{\rm H}} \times f_{\rm pol} (\lambda) F_{\rm turb} \sin^{2}\gamma,
\label{eq:Pmod_short}
\end{equation}
where
\begin{equation}
     \frac{P_{\rm i}(\lambda)}{N_{\rm H}} =  \int_{a_{\rm min}}^{a_{\rm max}} C_{\rm pol}(a,\lambda) n_{d}(a)da = \sigma_{\rm pol, i}(\lambda),
\label{eq:Pi}
\end{equation}
is the intrinsic polarization efficiency, corresponding to the intrinsic polarization coefficient $\sigma_{\rm pol, i}(\lambda)$ for all grain size distribution having perfect magnetic alignment. This parameter depends only on the intrinsic properties of dust grains (i.e., grain size distribution and elongation). Among that,
\begin{equation}
f_{\rm pol} (\lambda)  = \frac{\sigma_{\rm pol, align}(\lambda)}{\sigma_{\rm pol, i}(\lambda)},
\label{eq:fpol}
\end{equation}
is the fraction of polarization coefficient describing how efficiently aligned grains can polarize background sources (see \citealt{TruongHoang2025}). Lower $f_{\rm pol}(\lambda)$ corresponds to inefficient alignment with B-fields contributing to lower starlight polarization efficiency. The determination of $f_{\rm pol}(\lambda)$ strongly depends on the alignment function $f_{\rm align}(a)$ in local environments, which can be characterized based on the MRAT alignment theory (\citealt{Hoang2016a}, see Section \ref{sec:falign_RAT} and Appendix \ref{sec:appendix_RAT}). 



From Equation \ref{eq:Pmod_short}, the inferred inclination angles can be derived from the observed starlight polarization efficiency $P_{\lambda}/N_{\rm H}$ as
\begin{equation}
    \sin^{2}\gamma = \frac{1}{f_{\rm pol}(\lambda)F_{\rm turb}}\frac{P_{\lambda}/N_{\rm H}}{P_{\rm i}(\lambda)/N_{\rm H}},\label{eq:chi_ext}
\end{equation}
which provides the absolute values of $|\gamma|$ (i.e., the mean orientation of B-fields with respect to the LOS) when $\sin^2\gamma \lesssim 1$. 

Equations (\ref{eq:chi_ext}) combined with (\ref{eq:Pi}) and (\ref{eq:fpol}) reveal that, to derive the inclination angles of B-fields, we need to know the following parameters:
\begin{itemize}
    \item observed polarization efficiency, $P_{\lambda}/N_{\rm H}$;
    \item depolarization caused by magnetic tangling, $F_{\rm turb}$; 
    \item dust properties (optical constant, size distribution, grain shape);
    \item grain alignment function, $f_{\rm align}(a)$.
\end{itemize}

The first two parameters can be constrained from starlight polarization observations, whereas the latter two require a combination of grain-alignment physics and observations of both starlight polarization and extinction. In the following sections, we will describe how these parameters are determined from observations and the alignment theory.



\section{Observations of Starlight Polarization}
\label{sec:obs}
In this section, we will describe the archival starlight polarization data by \cite{Chen2023} and present the derivation of the observed polarization efficiency and the magnetic turbulence factor $F_{\rm turb}$ from the available polarimetric observation.   

\subsection{$K_s$-band Polarization from ISRF/SIRPOL}
To obtain the observed starlight polarization efficiency toward G11, we use the archival data of polarimetric observations from background stars by \cite{Chen2023}. The data was taken by SIRPOL - a NIR polarimeter on the 1.4 m InfraRed Survey Facility (IRSF) telescope at the South Africa Astronomical Observatory (SAAO)(see, e.g., \citealt{Nagayama2003}; \citealt{Kandori2006}). The instrument has the capability of polarimetric imaging simultaneously at $J$, $H$ and $K_s$-bands (e.g., $\lambda_{\rm J} =1.22\,\rm\mu m$, $\lambda_{\rm H} = 1.63\,\rm\mu m$ and $\lambda_{\rm K_s} = 2.19\,\rm\mu m$) over the field-of-view of $7.7 ' \times 7.7 '$. \cite{Chen2023} observed a mosaic of three SIRPOL frames positioning around G11 during the night of 2018 July 21. The data reduction of polarization observations was carried out by \cite{Chen2023}. 




\begin{figure}
    \includegraphics[width=1.0\linewidth]{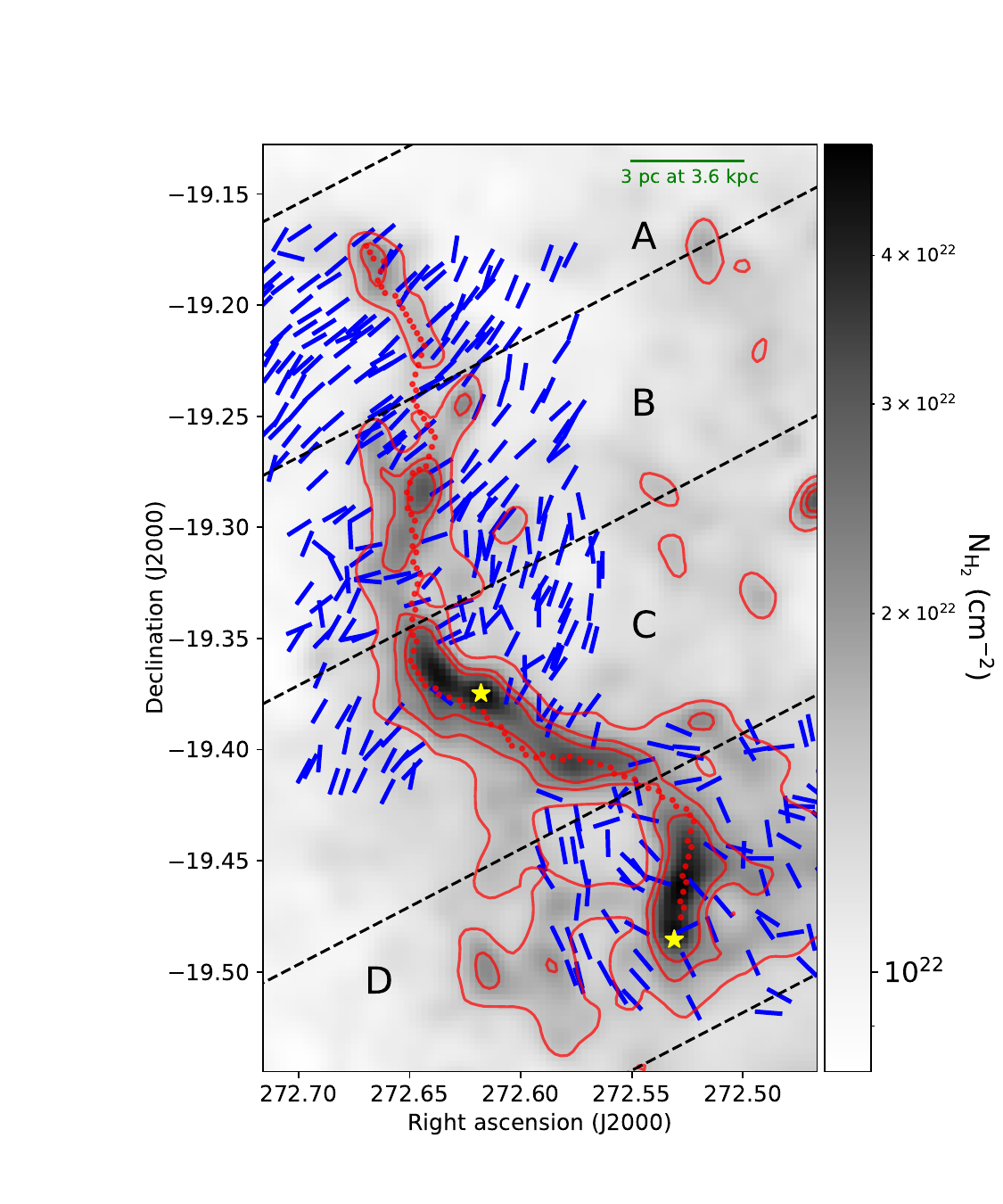}
    \caption{The foreground-corrected $K_{s}$-band polarization with high signal-to-noise $P_{\rm K}/\delta P_{\rm K} > 3$ measured by SIRPOL at $2.19\,\rm\mu m$ (blue segments, see \citealt{Chen2023}). The orientation of polarization indicates the orientation of POS B-fields in the outer regions of G11. The length of the blue segments is in an arbitrary scale. The polarization vectors are overplotted onto a gray map of molecular hydrogen column density $N_{\rm H_2}$ derived from Herschel observations (\citealt{Zucker2018_filament}). The massive clumps P1 and P6 are marked by yellow stars (\citealt{Henning2010}). The red contours indicate the column density at 1.5, 2, and 2.5 $\times 10^{22}\,\rm\cm^{-2}$. The spine structure of the G11 filament is indicated by red circles. The G11 filament is divided into four sub-regions (Regions A, B, C, and D) along the spine in Galactic coordinates, whose boundaries are represented by dashed black lines.}
    \label{fig:Pol_Ks}
\end{figure}


The analysis of reddened background stars was carried out by combining the polarization observations with the $JHK_s$ photometry from the Vista Variables in the Via Lactea (a.k.a., VVVX) survey (see \citealt{Chen2022}; \citealt{Chen2023}). Based on the criteria toward G11 in the NIR color-color diagram (see in \citealt{Chen2023}), 364 stars observed at $K_{s}$ bands were classified as background stars, which are likely arising from dust grains aligned with B-fields within G11. The correction of foreground polarization contamination is carried out on the $K_{s}$-band polarization. We use directly the foreground-corrected $K_{s}$-band polarization degree $P_{K}(\%)$ with high signal-to-noise ratio (SNR) $P_{\rm K}/\delta P_{\rm K} > 3$ and polarization angle $\theta_{PA}$ analyzed by \cite{Chen2023}, which was publicly available on the Science Data Bank (\citealt{Chen_data}). The orientation of $K_{s}$-band polarization is illustrated in Figure \ref{fig:Pol_Ks} (blue segments). This also demonstrates the orientation of $B_{\rm POS}$ mostly in the outer regions of G11 observed by SIRPOL, in comparison to those traced by FIR polarization by SOFIA/HAWC+ in the inner dense filament by \cite{Ngoc2023}.

\subsection{Observed Polarization Efficiency}
We determine the starlight polarization efficiency at $K_s$-band, $P_{\rm K}/N_{\rm H}$. We consider the G11 filament is fully molecular, in which $N_{\rm H} \sim 2N_{\rm H_2}$ (\citealt{Chen2023}; \citealt{Ngoc2023}). The column density of molecular hydrogen $N_{\rm H_2}$ is derived from the modified black-body fitting to the Herschel multi-wavelength observations at 160, 250, 350, and 500 $\rm\mu m$ by \cite{Zucker2018_filament}. The resolution of the $N_{\rm H_2}$ map is $43''$ ($\sim 1.4\,\rm pc$) with a pixel size of $11''.5 \times 11''.5$, as shown in Figure \ref{fig:Pol_Ks}. The filamentary spine structure was determined using the \textit{RadFill} algorithm (see \citealt{Zucker2018} for further details). The spine structure of G11 is demonstrated by red points in Figure \ref{fig:Pol_Ks}.

Following by \cite{Chen2023}, we divide the starlight polarimetric observations into four sub-regions based on the gradient of the column density $N_{\rm H_2}$ along the filamentary spine in the Galactic coordinate: Region A ($11.^{\circ}24 < l < 11.^{\circ}.34$); Region B ($11.^{\circ}15 < l < 11.^{\circ}.24$); Region C ($11.^{\circ}04 < l < 11.^{\circ}.15$)
and Region D ($10.^{\circ}93 < l < 11.^{\circ}.04$). The division of four sub-regions is demonstrated in Figure \ref{fig:Pol_Ks} with region boundaries in Galactic longitudes (dashed black lines). The orientation of local $B_{\rm POS}$ tends to be perpendicular to the filament spine in Regions A, B, and C, while it is parallel to the spine structure in Region D (see the analysis of \citealt{Chen2023}). Then, we calculate the polarization efficiency $P_{\rm K}/N_{\rm H}$ and examine its variation in each sub-region along the filament spine. 

\begin{figure}
    \centering
    \includegraphics[width=1.0\linewidth]{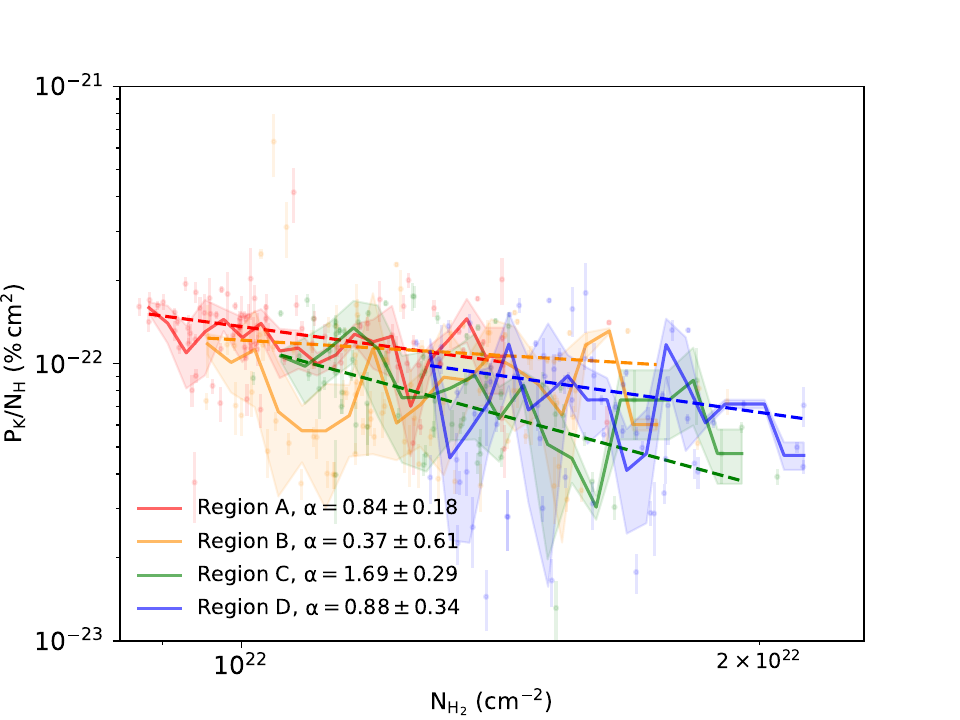}
    \caption{The starlight polarization efficiency $P_{\rm K}/N_{\rm H}$ vs. the molecular hydrogen column density $N_{\rm H_2}$ in four sub-regions along the spine of the G11 filament. Solid color lines demonstrates the running mean of the scattered data in each sub-region, while color shaded regions presents its $1\sigma$ uncertainty. A power-law fit with $P_{\rm K}/N_{\rm H} \varpropto N_{\rm H_2}^{-\alpha}$ is applied to the running mean of the observed data (dashed color lines).}
    \label{fig:PK_NH2}
\end{figure}

Figure \ref{fig:PK_NH2} illustrates the change in the observed polarization efficiency $P_{\rm K}/N_{\rm H}$ with respect to the local column density $N_{\rm H_2}$ in four sub-regions. The mean column density $\langle N_{\rm H_2}\rangle$, the mean polarization degree $\langle P_{\rm K} \rangle$ and the mean polarization efficiency $\langle P_{\rm K}/N_{\rm H} \rangle$ in each sub-region are summarized in Table \ref{tab:PK_NH2}. The observed data mainly resolve the POS B-fields in the outer regions up to $N_{\rm H_2} \sim 2 \times 10^{22}\,\cm^{-2}$. The running mean, weighted with the measurement uncertainty $\delta P_K/N_{H}$, is applied to demonstrate the local variations in the observed data (solid color lines). A power-law fit with $P_{\rm K}/N_{\rm H} \varpropto N_{\rm H_2}^{-\alpha}$ is applied (dashed color lines) to fit the weighted-running mean of the data for all sub-regions, with the slope $\alpha$ ranging from 0 to 1 along with its uncertainty. This slope characterizes the effectiveness of grain alignment: $\alpha = 1$ corresponds to a complete loss of grain alignment, while shallower $\alpha < 1$ shows the efficient alignment induced by RATs (see \citealt{Whittet2008}; \citealt{Hoang2021}). The lowest-density Region A tends to have the slope of $\alpha \sim 0.84$ and the highest mean polarization $\sim 1.28 \times 10^{-22}\,\%\,\rm cm^{-2}$. Region B has the shallowest slope ($\alpha \sim 0.3$); however, the data are highly scattered that causes the power-law fit to be less robust (i.e., having highest fitted slope uncertainty $\delta\alpha \gg \alpha$). This could be associated with the significant variations in 3D B-fields (i.e., 3D inclination angles and magnetic turbulence) across this filamentary region. Toward denser filamentary regions (Regions C and D), the fitted slope is steeper with $\alpha  > 0.8$ and the polarization efficiency is lower due to the grain alignment loss owing to increasing gas randomization and strong B-field tangling. In particular, for Region C, the observed data show a slope much larger than 1 with $\alpha \sim 1.7$. The much steeper slope in this region could be explained by the effect of the inclined B-fields with respect to the LOS.


\begin{table}[]
\centering
    \caption{Summary of the mean column density $\langle N_{\rm H_2} \rangle$, the mean polarization degree $\langle P_K \rangle$ and starlight polarization efficiency $\langle P_K / N_{\rm H} \rangle$ in four sub-regions along the G11 filament.}
    \begin{tabular}{l c c c}
    \toprule
      Region & $\langle N_{\rm H_2} \rangle$  & $\langle P_K \rangle$  &  $\langle P_K/ N_{\rm H} \rangle$ \\

      & ($\times 10^{22}\,\rm cm^{-2}$) & ($\%$) & ($\times 10^{-22}\,\%\,\rm cm^{2}$) \\

      \midrule

      A & $1.01$ & 2.7  & $1.28$\\

      B & $1.24$ & 2.2  & $0.8$\\

      C & $1.28$ & 2.3  & $0.85$\\

      D & $1.54$ & 2.3 & $0.7$\\
  
      \bottomrule
    \end{tabular}
    \label{tab:PK_NH2}
\end{table}

\subsection{Determining Magnetic Turbulence Factor from Starlight Polarization}
\label{sec:Fturb}
We obtain the magnetic turbulence factor $F_{\rm turb}$ from the dispersion of polarization angles derived from the observed starlight polarization observation. We assume the local magnetic turbulence in G11 is isotropic. Therefore, the effect of B-field fluctuation on the depolarization can be described from the POS polarization angle dispersion (\citealt{TruongHoang2025}). For the estimation of the polarization angle dispersion, we split the starlight polarization data into many $2 ' \times 2 '$ grid cells. This choice allows the stellar density within a cell to be sufficiently high for characterizing the magnetic turbulence effect (the number of polarized background stars $N_{\rm star} \gtrsim 3$, see Table 2 in \citealt{Chen2023}). The polarization angle dispersion for each cell, denoted by $\sigma_{\theta}$, is calculated by using the unsharp-masking method as (see \citealt{Pattle2017}; \citealt{Hwang2021})
\bea
\sigma_{\theta}=\left[\frac{1}{N_{\rm star}}\sum_{i=1}^{N_{\rm star}}\left(\theta_{\rm PA, i}-\bar{\theta}_{\rm PA}\right)^{2}\right]^{1/2},\label{eq:sigma_theta}
\ena
where
\bea
\bar{\theta}_{\rm PA} = \frac{1}{2}\arctan[\langle \sin{(2\theta_{\rm PA, i})} \rangle, \langle \cos{(2\theta_{\rm PA, i})} \rangle],
\ena
is the circular mean, which is derived from the average sine and cosine of the polarization angle components $2\theta_{\rm PA, i}$ within a grid cell as (\citealt{Valdivia2022})
\bea
\langle \sin(2\theta_{{\rm PA},i}) \rangle &=& \frac{1}{N_{\rm star}}\sum_{i=1}^{N_{\rm star}}\sin(2\theta_{{\rm PA},i}),\nonumber\\
\langle \cos(2\theta_{{\rm PA},i}) \rangle &=& \frac{1}{N_{\rm star}}\sum_{i=1}^{N_{\rm star}}\cos(2\theta_{{\rm PA},i}).
\ena
For each grid cell, the angle differences $\theta_{\rm PA, i}-\bar{\theta}_{\rm PA}$ are wrapped in a range of $[0^{\circ} - 90^{\circ}]$ to avoid the 180-degree ambiguity. The polarization angle dispersion can also be calculated from the second-order angular dispersion structure function (a.k.a. ADF method, see \citealt{Hilderband2009}; \citealt{Houde2009}), or circular standard deviation of the polarization angle distribution (\citealt{Pelgrims2021}; \citealt{Polychronakis2025}); but for this study, we use only the unsharp-masking method onto this starlight polarization dataset.

Polarization angle measurements have their uncertainties as $\theta_{\rm PA,i} \pm \delta\theta_{\rm PA,i}$ (\citealt{Chen2023}). For each star in a $2 ' \times 2 '$ grid cell, we perform Monte Carlo (MC) simulations by drawing the polarization angle followed by a Gaussian distribution with a mean of $\theta_{\rm PA,i}$ and a width of $\delta\theta_{\rm PA,i}$, calculating the dispersion angle as shown in Equation \ref{eq:sigma_theta}, and repeating the process 1000 times. We determine the mean and standard deviation of the dispersion angle in each grid cell, denoted by $\bar{\sigma}_{\theta}$ and $\delta \sigma_{\theta}$. We note that this is the upper limit of the polarization angle dispersion, as it can be overestimated by the measurement errors (see in \citealt{Pattle2017}). We choose the cells where the contribution of measurement uncertainties is minimal (i.e., $\bar{\sigma}_{\theta} \gg \delta \sigma_{\theta}$). The turbulence factor can be estimated locally from the mean dispersion $\bar{\sigma}_{\rm \theta}$ as $F_{\rm turb} \approx 1 - 1.5\sin^2(\bar{\sigma}_{\rm \theta})$.

\begin{figure*}
    \includegraphics[width=0.5\linewidth]{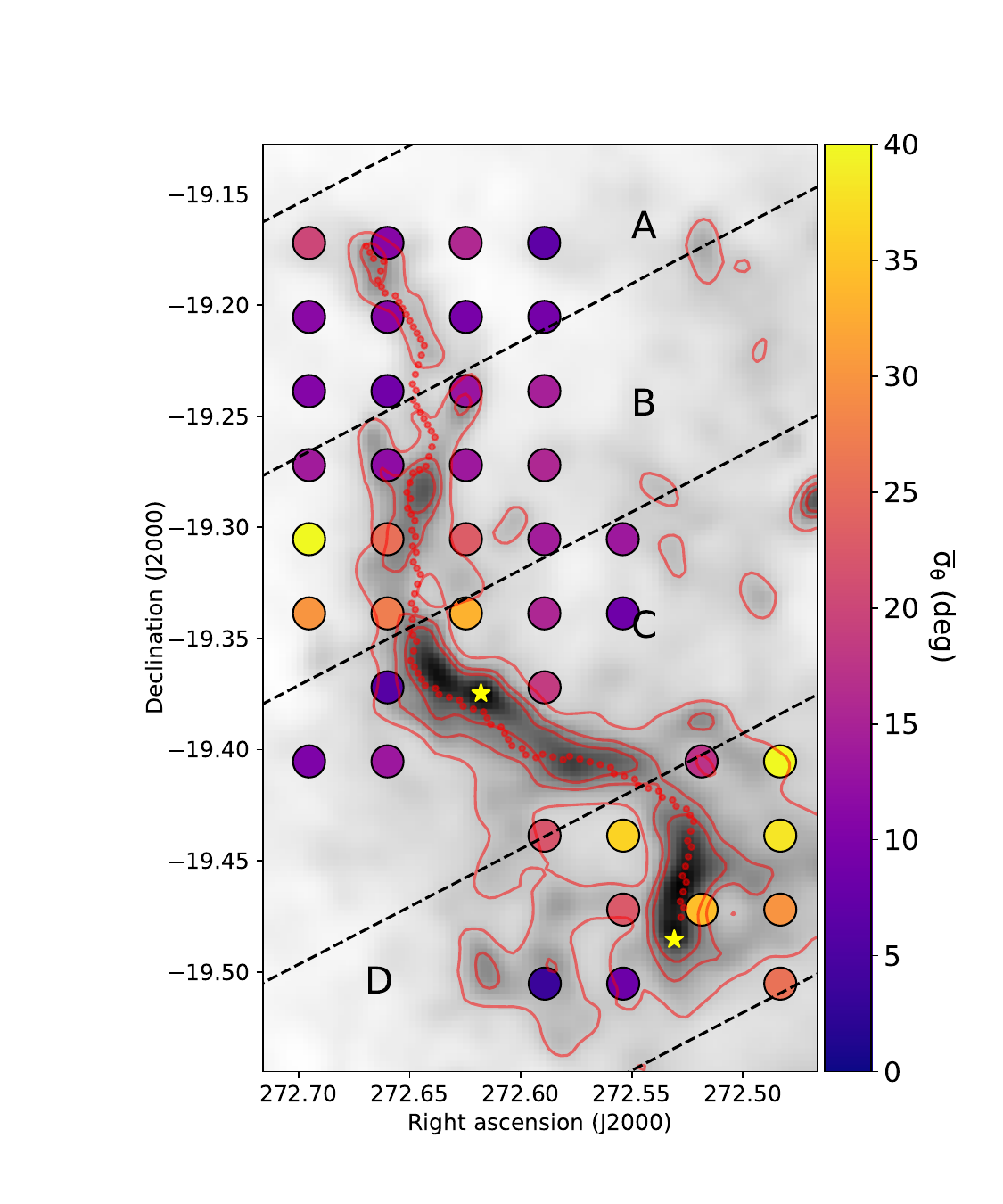}
    \includegraphics[width=0.5\linewidth]{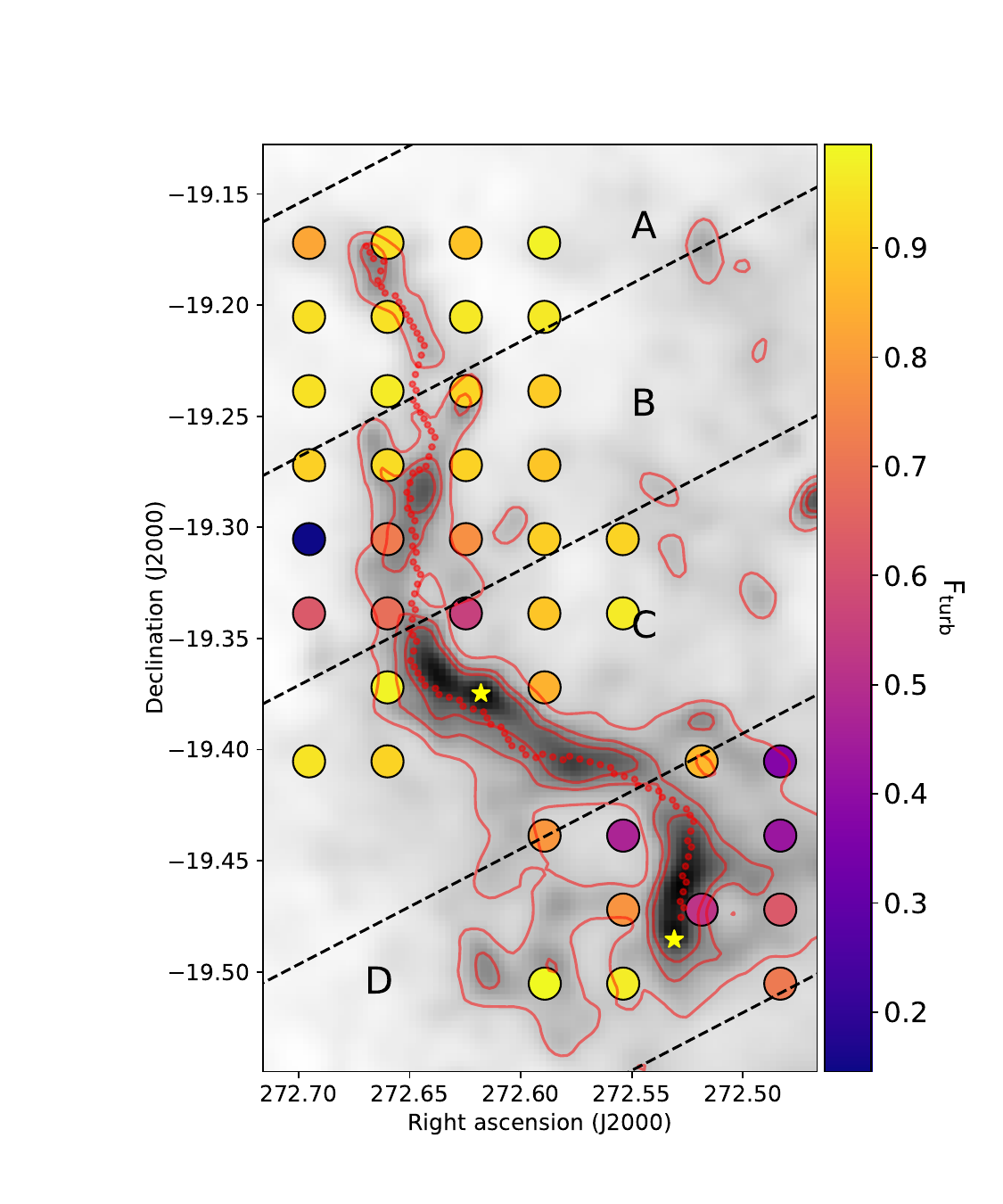}
    \caption{Left panel: The spatial distribution of the polarization angle dispersion $\bar{\sigma}_{\rm \theta}$ in each grid cell spacing of $2' \times 2 '$. Right panel: The local magnetic turbulence factor $F_{\rm turb}$ calculated from the polarization angle dispersion.}
    \label{fig:Fturb_Star}
\end{figure*}

Figure \ref{fig:Fturb_Star} shows the spatial distributions of the mean POS polarization angle dispersion $\bar{\sigma}_{\rm \theta}$ (left panel) and the magnetic turbulence factor $F_{\rm turb}$ (right panel) within the G11 filament. In the outer regions of G11, the effect of magnetic turbulence is less dominant with lower $\bar{\sigma}_{\rm \theta} \lesssim 15^{\circ}$ and $F_{\rm turb} \sim 0.8 - 0.9$. The magnetic fluctuation effect on starlight polarization becomes significant in denser filaments with $\bar{\sigma}_{\rm \theta} \sim 20^{\circ} - 50^{\circ}$, leading to lower $F_{\rm turb} < 0.5$.

\section{Dust Properties and Alignment Function}
\label{sec:dust_align}
In this section, we present the characterization of the intrinsic dust properties from archival starlight polarization and interstellar dust extinction observations. We present the grain alignment physics in G11 based on RAT theory and the derivation of the intrinsic polarization efficiency and the polarization coefficient when the dust and grain alignment properties are given.

\subsection{Dust Properties}
\subsubsection{Dust model and grain size distribution}
\label{sec:int_amax}

For the dust model, we consider a mixture of Astrodust (i.e., silicate and carbonaceous components are mixed into a single grain; see \citealt{Draine2021}) and polycyclic aromatic hydrocarbons (PAHs) presented in nanosized grains (\citealt{Leger1984}; \citealt{Allamandola1985}; \citealt{Draine2021_PAH}). The Astrodust + PAHs model was applied to explain the extinction, emission, and polarization observations in the diffuse ISM by \cite{Hensley2023}.  We consider the grains to be oblate spheroids with an axial ratio of $s =b/a>1$ where $b$ and $a$ are the lengths of the semi-major and semi-minor axis. In this study, we assume that the grain elongation does not significantly change from the outer to inner regions of the filament. However, a theoretical study suggests that the grain elongation can increase with grain growth due to the effect of grain alignment (\citealt{Hoang2022_Graingrowth}).

We adopt the combination of the Astrodust and PAH size distributions. The size distribution of Astrodust is considered to follow the Mathis-
Rumpl-Nordsieck (MRN) power-law distribution with $n_{\rm Astro}(a) \varpropto a^{-3.5}$(\citealt{Mathis1977}). The assumption of a power-law distribution is reasonable for grains grown in dense filamentary cloud and starless cores due to accretion and coagulation (\citealt{Hirashita2012}; \citealt{Bate2022}). The PAH size distribution is considered to follow the two log-normal size distributions described by \cite{DraineLi2007} found in the ISM (see in Appendix \ref{sec:appendix_GSD}). The grain sizes cover within the range from the minimum size $a_{\rm min} = 3.5\,\rm \AA$ to the maximum size $a_{\rm max}$, where $a_{\rm max}$ is the constrained value from observations. The value of $a_{\rm min} = 3.5\,\rm \AA$ is chosen as the lower limit of grain size being determined by thermal sublimation (\citealt{Draine1979}; \citealt{DraineLi2007}). The values of $a_{\rm max}$ could increase with increasing gas density owing to the grain growth effect (see \citealt{Hirashita2012}; \citealt{Bate2022}). 

\subsubsection{The maximum size from dust extinction curve}
To determine the maximum grain size $a_{\rm max}$ in G11, we use the low-resolution archival data of the interstellar dust extinction curve within the Milky Way by \cite{Zhang2025}. The authors used the mean blue and mean red photometer spectra for 130 million stars provided by the Gaia mission (\citealt{Gaia2023}; \citealt{Montegriffo2023}) and applied a forward machine learning model to predict the stellar model and dust extinction curve from optical to NIR wavelengths for each star along the LOS (see \citealt{Zhang2023}). The shape of the extinction curve is characterized by the total-to-selective extinction ratio, $R_V$, as 
\begin{equation}
    R_V = \frac{A_V}{A_B - A_V},
\label{eq:Rv}
\end{equation} 
where $A_B$ and $A_V$ are the extinction magnitude at B and V bands ($\lambda_{\rm B} = 0.45\,\rm\mu m$ and $\lambda_{\rm V} = 0.55\,\rm\mu m$). The total-to-selective ratio $R_V$ was generated from the observed Gaia data in a 3D space with respect to angular positions in the Galactic coordinate and distances up to a few kiloparsecs in the Milky Way, and is publicly available (\citealt{Zhang2025}). 

\begin{figure}
    \includegraphics[width=1.0\linewidth]{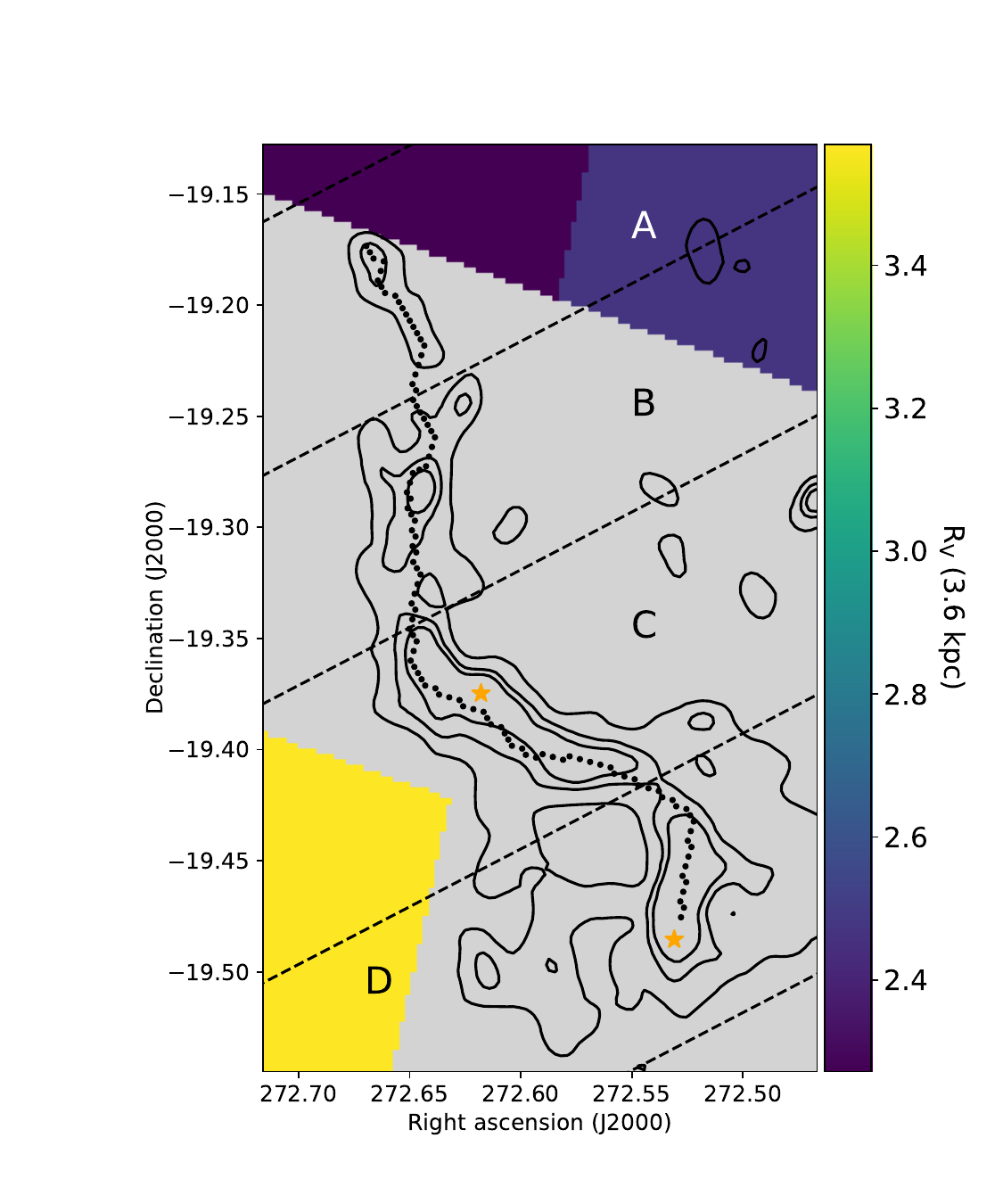}
    \caption{The spatial distribution of the total-to-selective extinction ratio $R_{\rm V}$ in the outermost region of G11 derived from interstellar dust extinction data from background stars provided by the Gaia mission (\citealt{Zhang2025}). The lightgray color presents the NAN values of $R_{\rm V}$ due to the non-detection of background stars beyond G11 at optical wavelengths by the Gaia mission.}
    \label{fig:Rv_obs}
\end{figure}

We use the differential $R_{55} = \Delta A_{55}/(\Delta A_{45} -\Delta A_{55})$ at the G11 distance of 3.6 kpc (\citealt{Pillai2016}) within the distance bin of 5 pc, where $\Delta A_{45}$ and $\Delta A_{55}$ are the differential extinction magnitudes at 450 and 550 nm, respectively (see in \citealt{Zhang2024, Zhang2025}). We take the differential $R_{55}$ centered at the position of G11 ($l = 11.^{\circ}119$ and $b = -0.^\circ 0647$, \citealt{Wang2014}). The observed optical $R_{\rm V}$ is converted from the differential $R_{55}$ by the empirical interstellar extinction law as $R_{\rm V} = 1.1R_{55} + 0.07$ (\citealt{Fitzpatrick2019}; \citealt{Zhang2024}). Figure \ref{fig:Rv_obs} shows the spatial distribution of $R_{\rm V}$ from the optical dust extinction curve (\citealt{Zhang2025}). The optical $R_{\rm V}$ mostly trace the dust properties in the outermost regions of G11 with $N_{\rm H_2} \lesssim 10^{22}\,\rm cm^{-2}$ (Figure \ref{fig:Pol_Ks}), ranging from 2.2 - 3.5. This is caused by the absorption and scattering of dust in the outer regions of G11. In the inner dense cloud with $N_{\rm H_2} > 10^{22}\,\rm cm^{-2}$, there are no background stars behind G11 detected by Gaia due to strong dust attenuation at optical wavelengths (450 and 550 nm), resulting in NAN values of differential $R_{55}$ (and also $R_{\rm V}$) in this region.



The maximum grain size $a_{\rm max}$ could be probed from the shape of the dust extinction curve, parameterized by the total-to-selective extinction ratio $R_V$ from Equation \ref{eq:Rv} (see, e.g., \citealt{Mathis1977}; \citealt{Cardelli1989}; \citealt{Weingartner2001}). We perform numerical modeling of the dust extinction curve from optical to NIR wavelengths ($\lambda = 0.1 - 20 \,\rm\mu m$) using the latest version of the \textsc{DustPOL\_py} code (\citealt{Tram2025}, see Appendix \ref{sec:appendix_extinction}). We consider the variations of maximum grain size $a_{\rm max}$ and grain elongation. The modeled $R_V$ is then calculated from the modeled $A_B$ and $A_V$ for different dust properties (see Appendix \ref{sec:appendix_extinction}). We compare with the observed $R_{\rm V}$ from Gaia data by \cite{Zhang2025} to derive the constraint on the maximum size in G11.

\begin{figure}
    \includegraphics[width=1.0\linewidth]{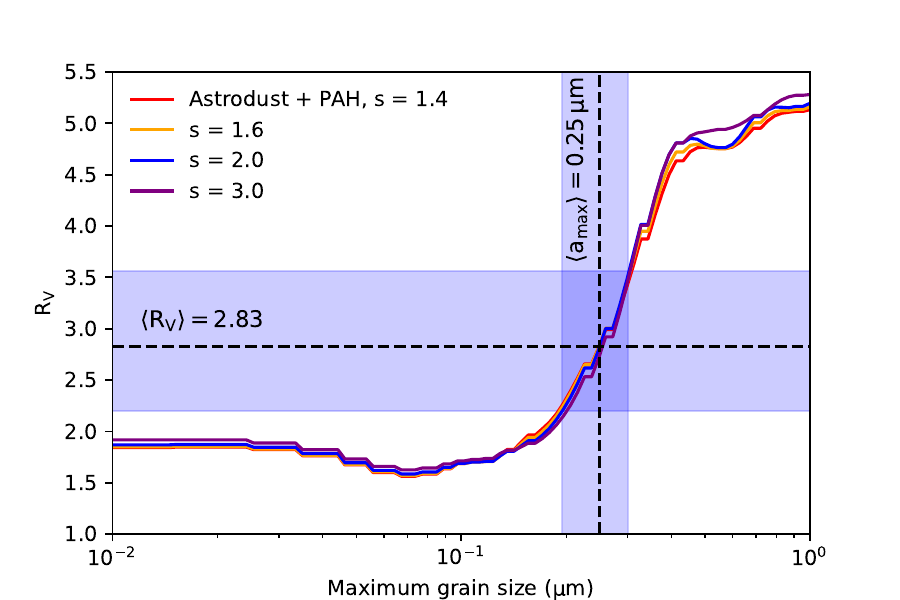}
    \caption{The variation of the modeled $R_{\rm V}$ generated by \textsc{DustPOL\_py} for Astrodust+PAHs grains, considering varying maximum grain size $a_{\rm max}$ and grain axial ratio of oblate spheroids $s > 1$. The $R_{\rm V}$ increases with increasing $a_{\rm max}$. The modeled $R_{\rm V}$ is best-fitted to the observed averaged optical $\langle R_{\rm V} \rangle \approx 2.83$ when the mean $\langle a_{\rm max} \rangle = 0.25\,\rm\mu m$.}
    \label{fig:Rv}
\end{figure}

Figure \ref{fig:Rv} demonstrates the dependence of the modeled $R_V$ on the maximum grain size $a_{\rm max}$ and grain axial ratios ($s>1$) for Astrodust+PAHs grains, adopting the size distributions described in Section \ref{sec:int_amax} (see Appendix \ref{sec:appendix_GSD}). The $R_V$ characterizes the upper limit of grain size: a larger $R_V$ indicates the presence of large grain sizes in local environments. From the observed averaged $R_V \sim 2.2 - 3.5$ toward the outermost region of G11 (Figure \ref{fig:Rv_obs}), we constrain $a_{\rm max} \sim 0.2 - 0.3 \mum $ (blue shaded region) possibly presented in the outermost region of the G11 filament. For this study, we use the observed averaged optical $\langle R_{\rm V} \rangle \sim 2.83$ for the G11 filament. This gives the mean maximum grain size $\langle a_{\rm max}\rangle = 0.25\,\rm\mu m$ (dashed black line), and is considered to be constant across G11. 



It is worth noting that the dust extinction curve is independent of the grain elongation, as demonstrated in Figure \ref{fig:Rv} for increasing grain axial ratio (color solid lines). The elongation of grains can be determined from the maximum observed polarization efficiency, and will be presented in the next Section \ref{sec:int_elongation}.

\subsection{Alignment Function from the RAT Theory}
\label{sec:falign_RAT}
This section describes in detail how the alignment function is calculated for G11 using the RAT theory.

\subsubsection{A model for the alignment function}
As described in Section \ref{sec:method}, to infer the inclination angles of the mean fields from starlight polarization efficiency, the alignment function $f_{\rm align}(a)$ must first be quantified. 

According to the RAT theory, the alignment function can be described by (\citealt{Hoang2016a,Lee2020})
\bea 
f_{\rm align}(a) = R\times \left[1 - \exp\left(-\left(\frac{a}{2a_{\rm align}}\right)^{3}\right)\right],
\label{eq:falign}
\ena
where $a_{\rm align}$ is the minimum size of grains that can be aligned by RATs (see Appendix \ref{sec:appendix_RAT}), and $R$ is the Rayleigh reduction factor characterizing the alignment degree of grains with the ambient B-fields. In the classical RAT theory, large grains $a > a_{\rm align}$ can be perfectly aligned with B-fields with $R = 1$. The exact value $R$ depends on the grain size and magnetic properties of grains according to the MRAT alignment theory (see \citealt{Hoang2016a}), and $R$ can be up to $1$ for superparamagnetic grains (SPM) that have embedded iron clusters (see Figure 2 of the alignment function for SPM grains in \citealt{TruongHoang2025}). 

As shown in Equation \ref{eq:align_ana}, the alignment size can be calculated when the environmental properties of G11 (i.e., gas density, gas and dust temperature) are provided.

\subsubsection{Volume Density and Dust Temperature from Herschel Multi-wavelength Observations}
\label{sec:nH2_Td}

To determine the gas volume density, we use the archival map of the molecular hydrogen column density $N_{\rm H_2}$ (see Figure \ref{fig:Pol_Ks}) and the dust temperature $T_{d}$ toward G11 derived from the Herschel multi-wavelength observations (\citealt{Zucker2018_filament}). The large-scale filamentary cloud is assumed to have a cylindrical shape, and the filament depth is equal to the filament width. The volume density of molecular hydrogen, denoted by $n_{\rm H_2}$, can be obtained from the column density $N_{\rm H_2}$ as
\begin{equation}
    n_{\rm H_2} = \frac{N_{\rm H_2}}{W},
\end{equation}
where $W$ is the width of the filament. The study of \cite{Kainulainen2013} from the combination of NIR and mid-IR (MIR) dust extinction by Spitzer and dust emission by Herschel found the extended self-gravitating cylindrical structure of G11 up to $\sim 5$ pc (see also \citealt{Chen2023}), giving $W = 5\,\rm pc$. 

\begin{figure*}
    \includegraphics[width=0.5\linewidth]{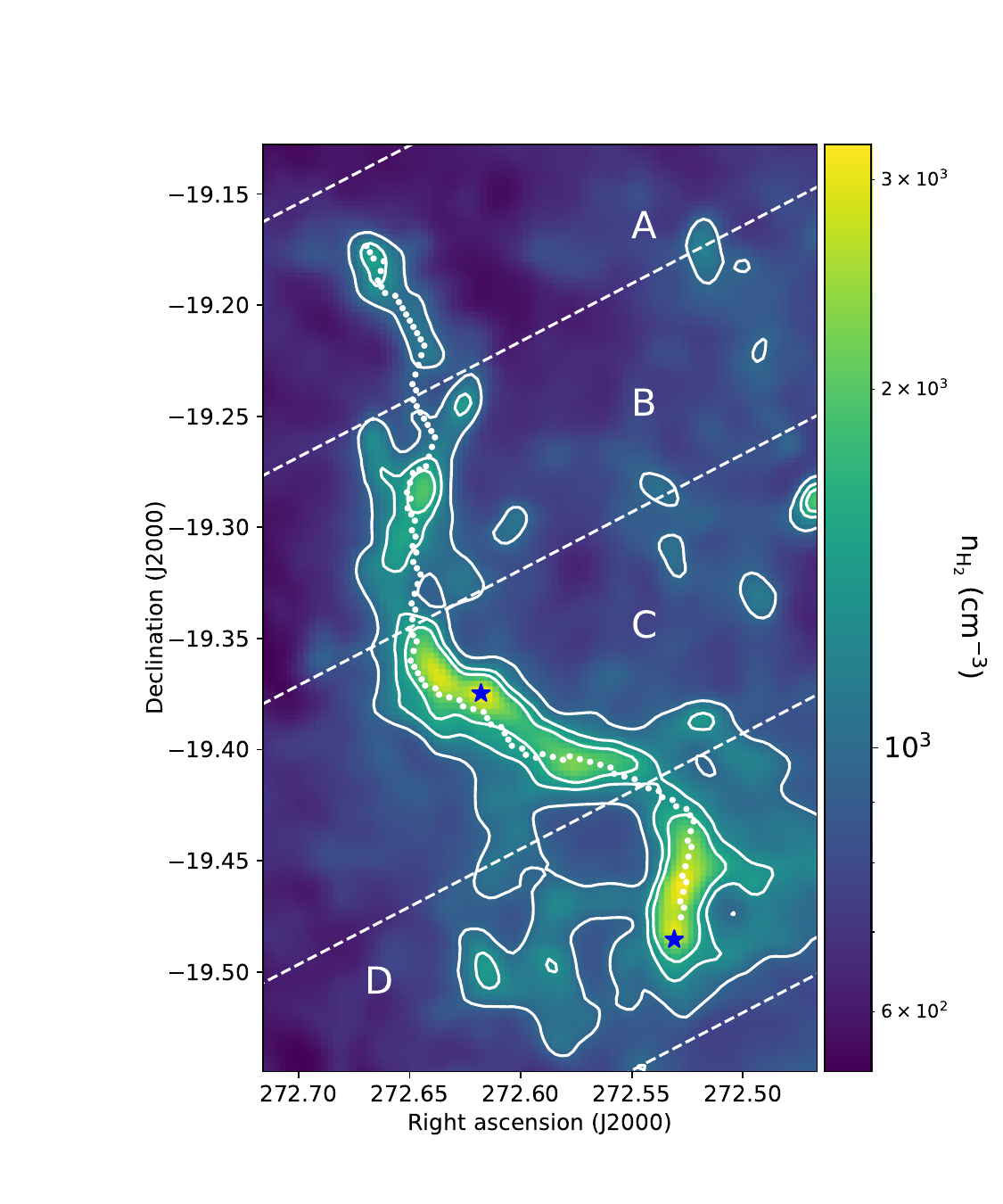}
    \includegraphics[width=0.5\linewidth]{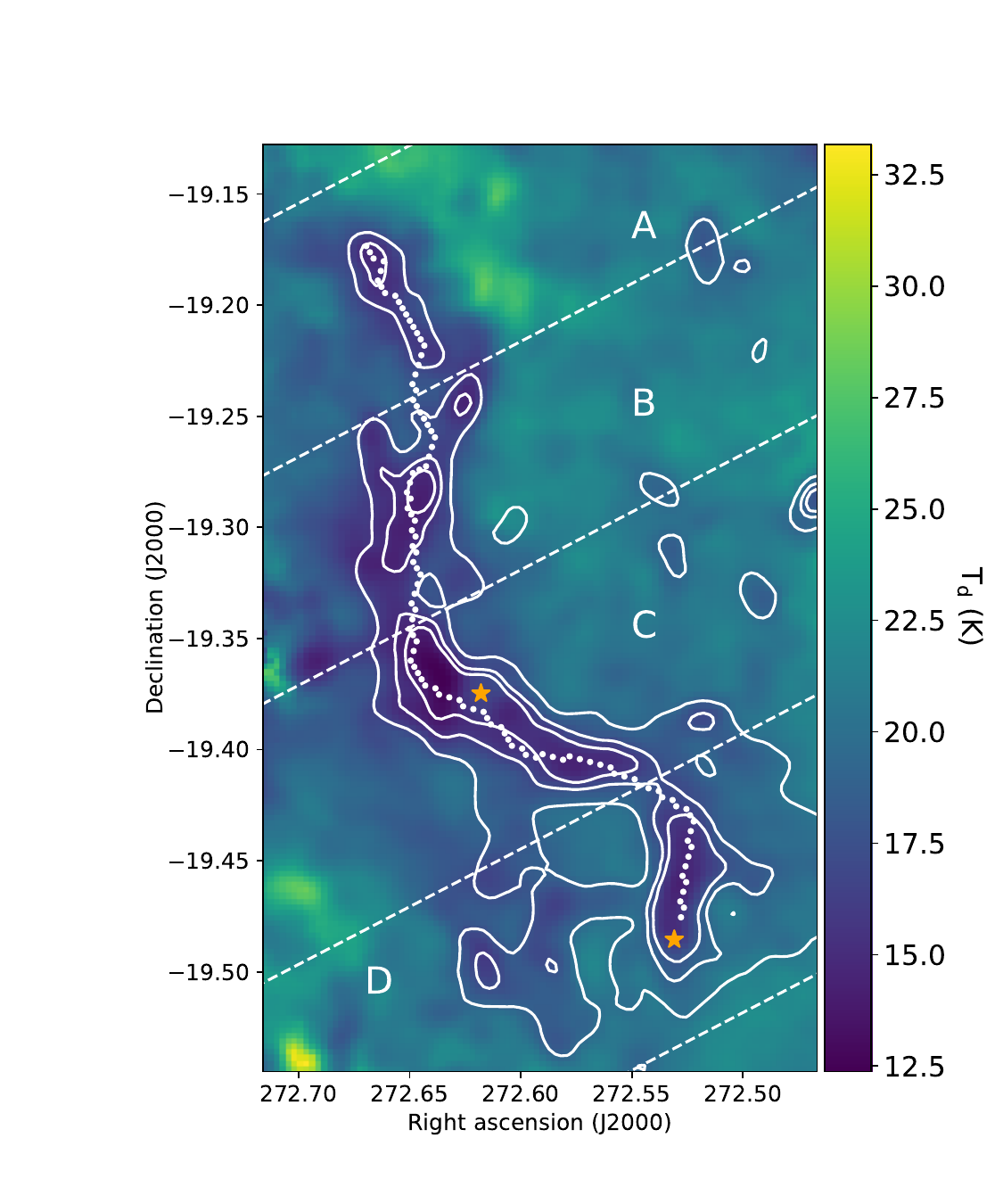}
    \caption{Maps of the volume density $ n_{\rm H_2}$ (left panel) and the dust temperature $T_{\rm d}$ (right panel) derived from multi-wavelength Herschel observations at 160, 250, 350, and 500 $\rm\mu m$ (see \citealt{Zucker2018_filament}).}
    \label{fig:nH2_Td}
\end{figure*}

The dust temperature $T_{\rm d}$ is taken from the modified black-body fitting to the Herschel multi-wavelength observations (\citealt{Zucker2018_filament}), with the same resolution as the $N_{\rm H_2}$ map. Figure \ref{fig:nH2_Td} shows the maps of the volume density $n_{\rm H_2}$ (left panel) and the dust temperature $T_d$ (right panel) in G11. The volume density of molecular hydrogen gas is $\sim 600 - 800\,\rm\cm^{-3}$ in the outer regions with $T_{\d} \sim 20 - 30\,\rm K$, and tends to increase toward the cold, denser parts with $n_{\rm H_2} \sim 2000 - 3000 \,\rm\cm^{-3}$ and $T_{\d} \sim 12 - 15\,\rm K$. It is noted that the derived volume density $n_{\rm H_2}$ is lower than that in the inner dense filament with a narrower width $W = 1\,\rm pc$, where the thermal dust polarization is mostly detected by SOFIA/HAWC+ (\citealt{Ngoc2023}). 

%


\subsubsection{Alignment Size and Alignment Function}
\label{sec:align}
To calculate the alignment size $a_{\rm align}$ for the dense G11 filament with no embedded bright sources, we consider the typical anisotropy degree of $\gamma_{\rm rad} = 0.1$ and the mean wavelength of $\bar{\lambda} = 1.2\,\rm\mu m$ for the diffuse interstellar radiation field. The gas is mainly heated by the grain-gas collision, giving the thermal equilibrium temperature $T_{\rm gas} = T_{\rm d}$ for dense and cold environments. With $T_{\rm d}$ and the hydrogen number density $n_{\rm H} = 2n_{\rm H_2}$ from Section \ref{sec:nH2_Td}, we consider the balance between the heating and cooling of silicate grains with sizes $a = 0.01 - 1\,\rm\mu m$ by the interstellar radiation field, giving the relationship between the radiation field strength and dust temperature as $U \simeq (T_{\rm d}/16.4\,\rm K)^{6}$ in the Rayleigh limit at long wavelengths with spectral index of $\beta = 2$ (\citealt{Draine2011book}).

Given the local environmental properties of G11 (i.e., $n_{\rm H}$ and $T_{\rm d}$), we can calculate the minimum size of aligned grains induced by RATs, $a_{\rm align}$, using the \textsc{DustPOL\_py} code (\citealt{Lee2020}; \citealt{Tram2021,Tram2024}, see Equation \ref{eq:align_ana} in Appendix \ref{sec:appendix_RAT}). 

Figure \ref{fig:align} shows the pixel-by-pixel map of the minimum aligned size in G11. The minimum aligned size is $\sim 0.03 - 0.05\,\rm\mu m$ in the outer regions with lower gas density $n_{\rm H_2} \lesssim 1000\,\cm^{-3}$, and increases to $a_{\rm align} \sim 0.12 - 0.15\,\rm\mu m$ in high-density regions as the grain alignment loss by strong gas randomization and the attenuation of interstellar radiation field is significant. 

We first assume perfect magnetic alignment of large grains $a > a_{\rm align}$ by RATs, which is considered the Ideal RAT realization with $R = 1$. When $a_{\rm align}$ and $R$ are provided, one can calculate the alignment function using Equation (\ref{eq:falign}). 

\begin{figure}
    \includegraphics[width=1.0\linewidth]{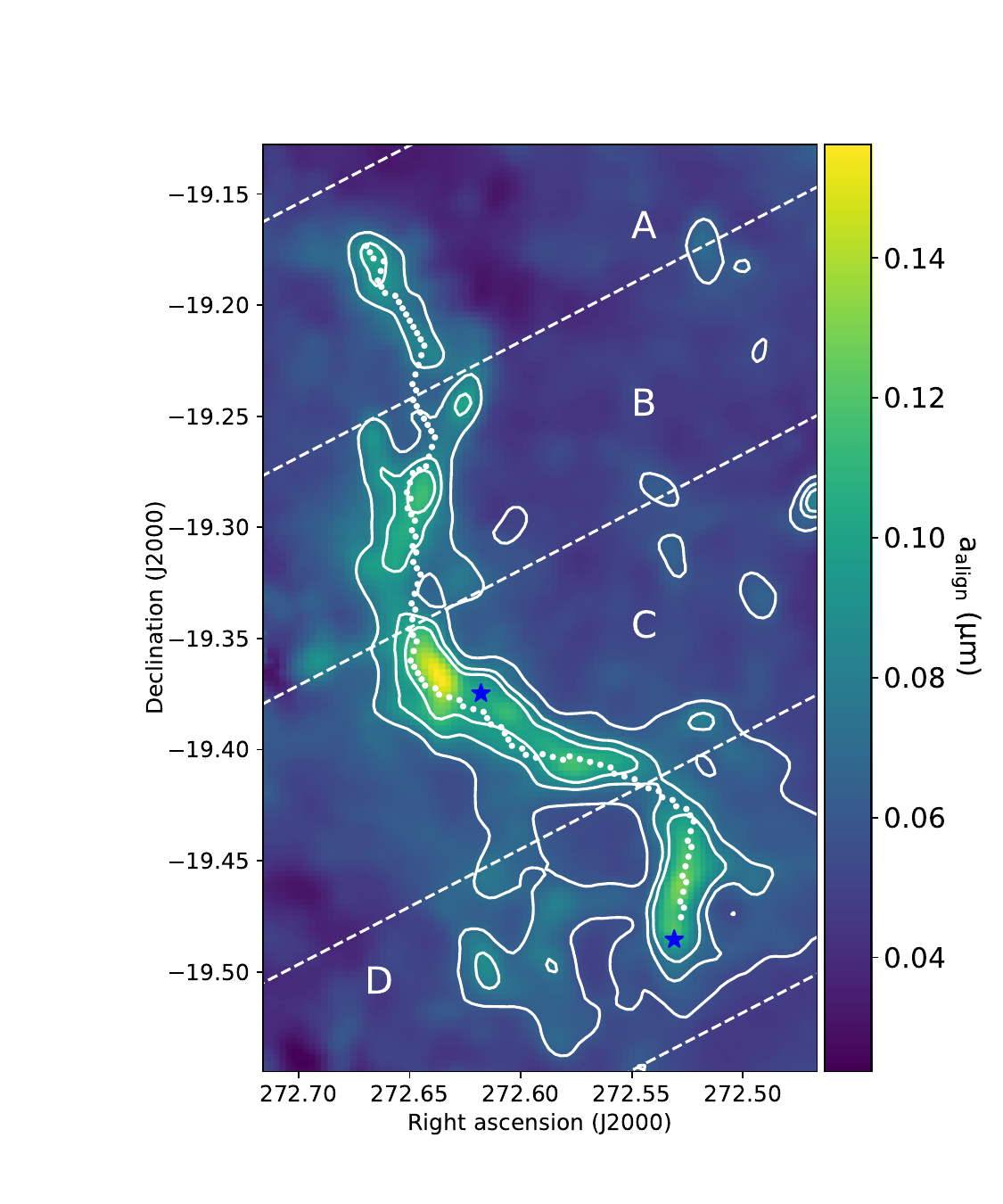}
    \caption{Map of the minimum aligned size $a_{\rm align}$ induced by RATs from interstellar radiation fields in the outer regions of G11.}
    \label{fig:align}
\end{figure}

\subsection{The Grain Axial Ratio or Grain Elongation}
\label{sec:int_elongation}
The shape of interstellar grains is a crucial parameter determining the polarization cross-section $C_{\rm pol}$ and the resulting polarization degree (\citealt{Draine2021}). However, this parameter is poorly constrained for interstellar dust. In this section, we will determine the axial ratio of the oblate grain shape (i.e., grain elongation) in the outermost regions of G11 by comparing the maximum observed starlight polarization efficiency with the modeled one derived from the \textsc{DustPOL\_py} code.

\begin{figure*}
    \includegraphics[width=0.5\linewidth]{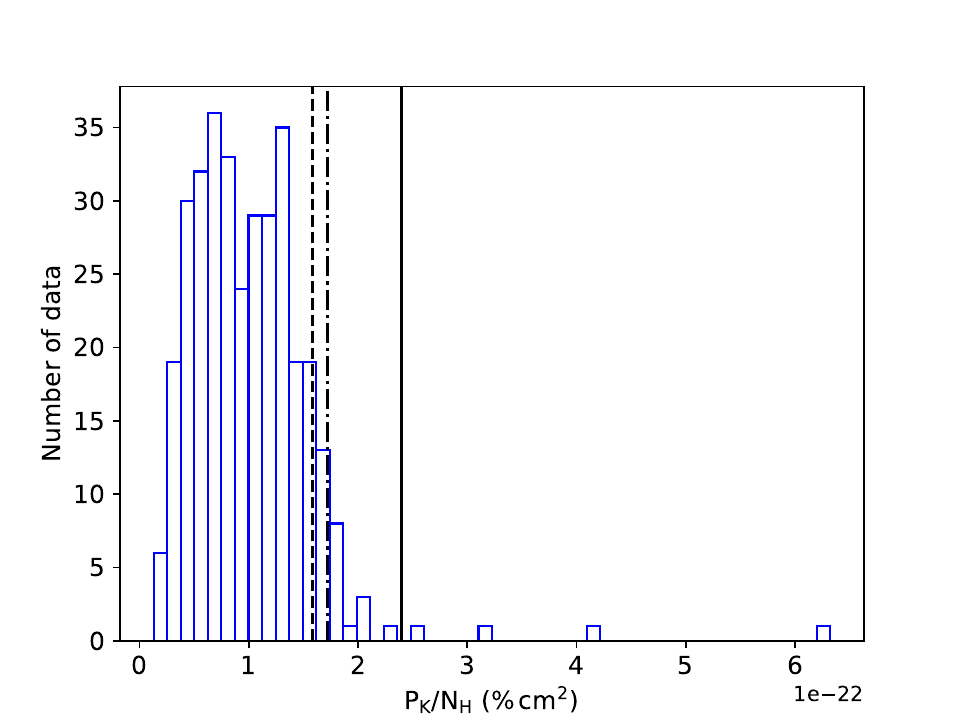}
    \includegraphics[width=0.5\linewidth]{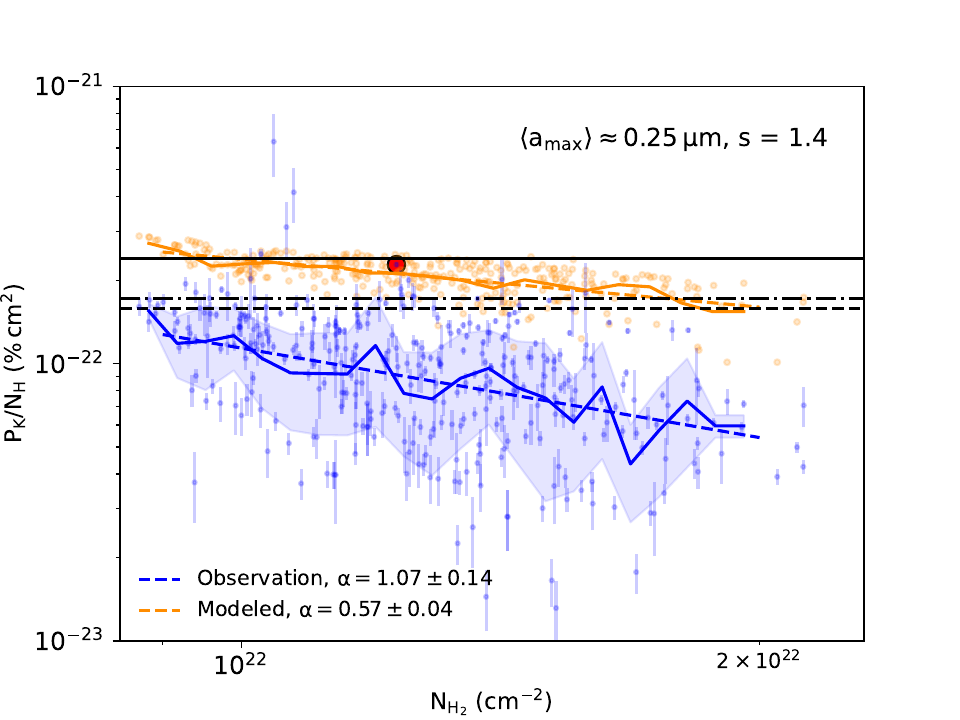}
    \caption{Left panel: Histogram of the observed starlight polarization efficiency $P_{\rm K}/N_{\rm H}$ in the G11 filament by \cite{Chen2023}. The 90th, 95th and 99th percentiles of the data are highlighted by dashed, dashed-dotted and solid black vertical lines, respectively. Right panel: The comparison between the observed and modeled polarization efficiency $(P_{\rm K}/N_{\rm H})_{\rm mod}$ by \textsc{DustPOL\_py} code in the Ideal conditions: (1) Ideal RAT alignment, (2) B-fields are lying in the POS and well-ordered (e.g., $\sin^2\gamma_{\rm obs} = 1$ and $F_{\rm turb} = 1$). The red mark point demonstrates the observed maximum polarization efficiency $(P_{\rm K}/N_{\rm H})_{\rm max}$. The modeled value is matched with the observed maximum value as the axial ratio of oblate grains of $s \gtrsim 1.4$.}
    \label{fig:PK_max}
\end{figure*}

The left panel of Figure \ref{fig:PK_max} shows the distribution of the observed $K_s$-band polarization efficiency $P_{\rm K}/N_{\rm H}$ in G11. The 90th, 95th and 99th percentiles of the data are demonstrated by dashed, dashed-dotted and solid black vertical lines, with the SNRs of $P_{\rm K}/\delta P_{\rm K} \sim 10, 5.5 $ and $56$, respectively. The upper bound of the polarization efficiency is set as the 99th percentile as $(P_{\rm K}/N_{\rm H})_{\rm max} \approx 2.3 \times 10^{-22}\,\%\,\cm^{2}$. The maximum value of polarization efficiency is located at $N_{\rm H_2} \approx 1.2 \times 10^{22}\,\rm\cm^{-2}$, where the grain alignment by RATs is efficient with $a_{\rm align} \sim 0.075\,\rm\mu m$ (see Figure \ref{fig:align}).  

We consider the ideal conditions of B-fields at the maximum polarization efficiency: (1) B-fields are parallel to the POS (i.e., $\sin^2\gamma = 1$) and (2) well-ordered (i.e., $F_{\rm turb} = 1$), and attribute the depolarization to the grain alignment and intrinsic dust properties. We consider the size distribution of Astrodust grains (see Appendix \ref{sec:appendix_GSD}) with fixed $\langle a_{\rm max}\rangle = 0.25\,\rm\mu m$ constrained from dust extinction (Section \ref{sec:int_amax}). We consider the variation of the grain axial ratio of $s = 1.4, 1.6, 2.0$ and 3.0 that is available in the Astrodust database. Using the polarization cross-section $C_{\rm pol}$ of Astrodust model with each grain elongation (\citealt{Draine2021}), and combining with the RAT alignment properties derived in Section \ref{sec:falign_RAT}, we use the \textsc{DustPOL\_py} code (\citealt{Tram2021,Tram2024}) to model the polarization efficiency at K-band, $(P_{\rm K}/N_{\rm H})_{\rm mod}$. The best-fit grain elongation can be obtained when $(P_{\rm K}/N_{\rm H})_{\rm mod} \approx (P_{\rm K}/N_{\rm H})_{\rm max}$ (\citealt{TruongHoang2025}).


The right panel of Figure \ref{fig:PK_max} presents the comparison between the observed profile $P_{\rm K}/N_{\rm H}$ vs. $N_{\rm H_2}$ (black) in G11 and the modeled profile numerically calculated by the \textsc{DustPOL\_py} code (red). The power-law fit with $P_{\rm K}/N_{\rm H} \varpropto N_{\rm H_2}^{-\alpha}$ is also performed on the running mean of both observed and modeled values. The slope of the modeled profile is shallower ($\alpha \sim 0.5$), in comparison to the observed one due to the effect of varying B-field geometries on the depolarization (i.e., varying $\sin^2\gamma$ and $F_{\rm turb}$). The observed maximum polarization efficiency is marked by a red point. The modeled maximum polarization efficiency is matched with the observed value when $s = 1.4$ - the same as the constrained lower value found in the diffuse ISM observations by \cite{Hensley2023}. Note that this is the lower limit of grain elongation that can be presented in the outer regions of G11 as grains could be more elongated (i.e., high $s$) in the denser filament by the growth of magnetically aligned grains (\citealt{Hoang2022_Graingrowth}).


\section{Intrinsic Polarization Efficiency and Polarization Coefficient Fraction at $K_s$-band}
\label{sec:int_fpol}
Giving the constrained $\langle a_{\rm max} \rangle = 0.25\,\rm\mu m$ and $s = 1.4$ from both the dust extinction curve at optical-NIR wavelengths and the maximum observed polarization efficiency presented in Sections \ref{sec:int_amax} and \ref{sec:int_elongation}, we can determine the intrinsic polarization efficiency following Equation \ref{eq:Pi}. The intrinsic polarization efficiency at $K_s$-band is $P_{i,K}/N_{\rm H} \approx 3.02 \times 10^{-22}\,\%\,\cm^{-2}$, and is considered to distribute uniformly in the G11 filament.

\begin{figure}
    \includegraphics[width=1.0\linewidth]{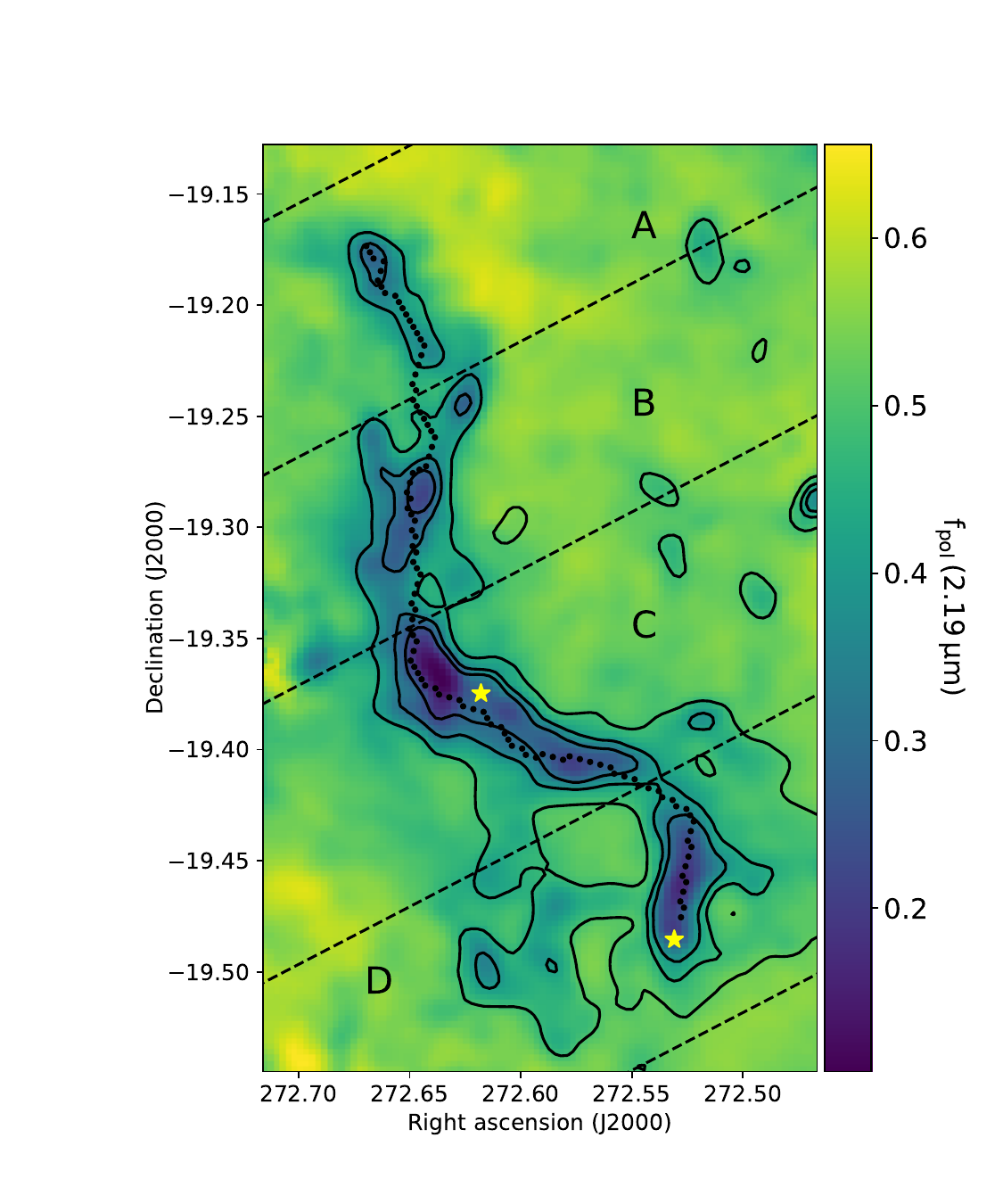}
    \caption{Map of the polarization coefficient fraction $f_{\rm pol, K}$ at $K_s$ band derived from the local RAT alignment properties and the constrained intrinsic dust properties in G11. The $f_{\rm pol,K}$ decreases toward the denser filament due to the grain alignment loss by increasing gas randomization.}
    \label{fig:fpol}
\end{figure}

When the intrinsic dust properties and alignment function are determined, we can calculate the polarization coefficient fraction $f_{\rm pol}$ at $K_s$-band using Equation \ref{eq:fpol}. Figure \ref{fig:fpol} shows the map of the polarization coefficient fraction $f_{\rm pol,K}$ in the G11 filament. The $f_{\rm pol}$ at $\lambda_{\rm K} = 2.19\,\rm\mu m$  is roughly 0.6 produced by the alignment of grain sizes $a < a_{\rm max} = 0.25\,\rm\mu m$ (see in \citealt{TruongHoang2025}). As the grain alignment loss is significant in the denser filament, the values of $f_{\rm pol, K}$ tend to decrease to $f_{\rm pol} < 0.3$.


\section{Results for B-Field Inclination Angles and 3D B-field Morphology}
\label{sec:result_incl}
In this section, we present the derivation of the inferred inclination angles of B-fields from starlight polarization efficiency using the technique of \cite{TruongHoang2025}. We present the construction of 3D B-field morphology when inferred angles are provided.

\label{sec:inferred_incl}
\subsection{Inferring B-field inclination angle}
Giving the intrinsic polarization efficiency from the constrained dust properties, the local impacts of grain alignment by RATs ($P_{\rm i,K}/N_{\rm H}$ and $f_{\rm pol, K}$, see Section \ref{sec:int_fpol}) and magnetic turbulence ($F_{\rm turb}$, see Section \ref{sec:Fturb}) within the G11 filament, we derive the local 3D inclination angles of the mean fields from the observed starlight polarization efficiency, $P_{\rm K}/N_{\rm H}$, as demonstrated in Equation \ref{eq:chi_ext}.

\begin{figure*}
    \includegraphics[width=0.48\linewidth]{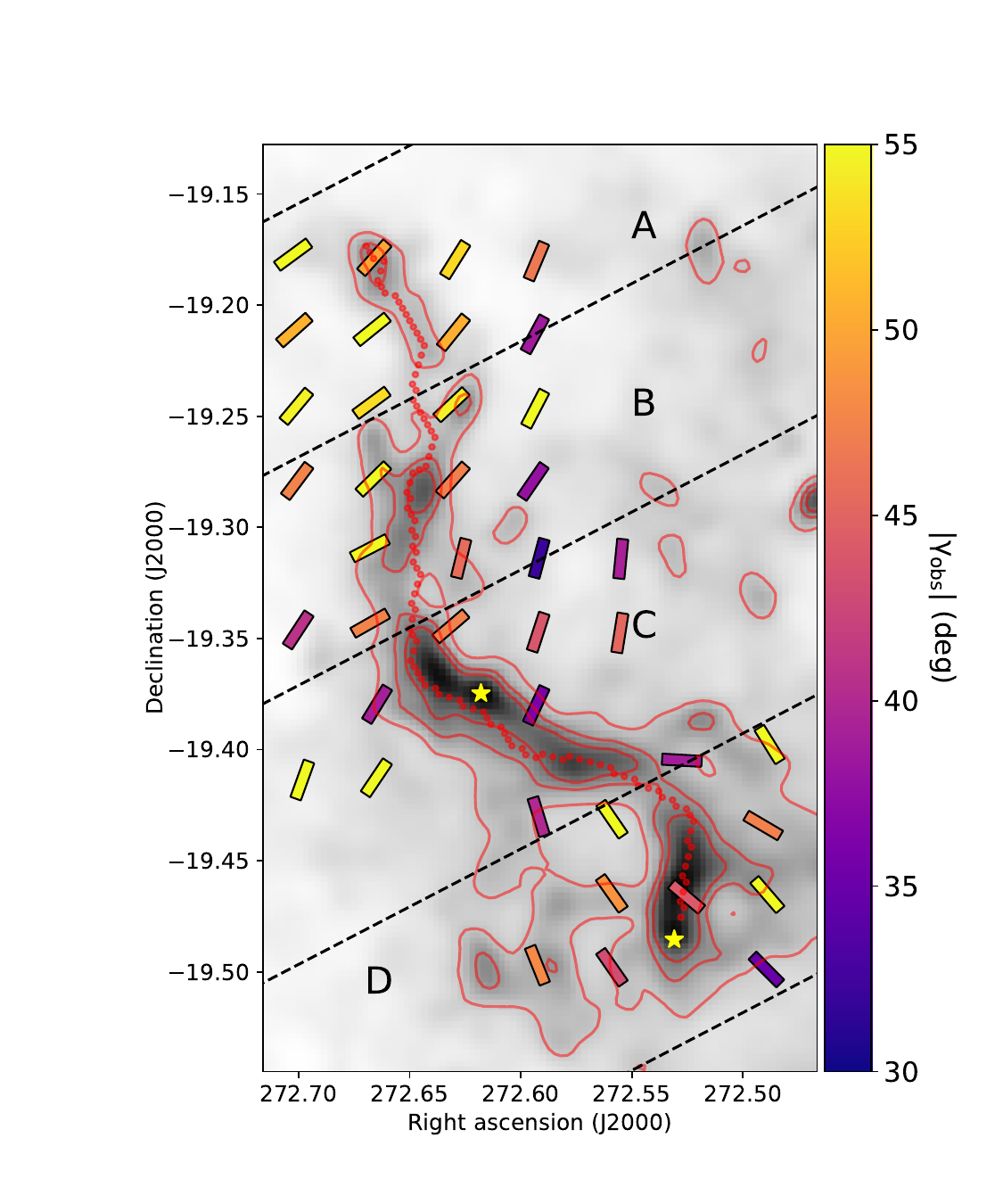}
    \includegraphics[width=0.48\linewidth]{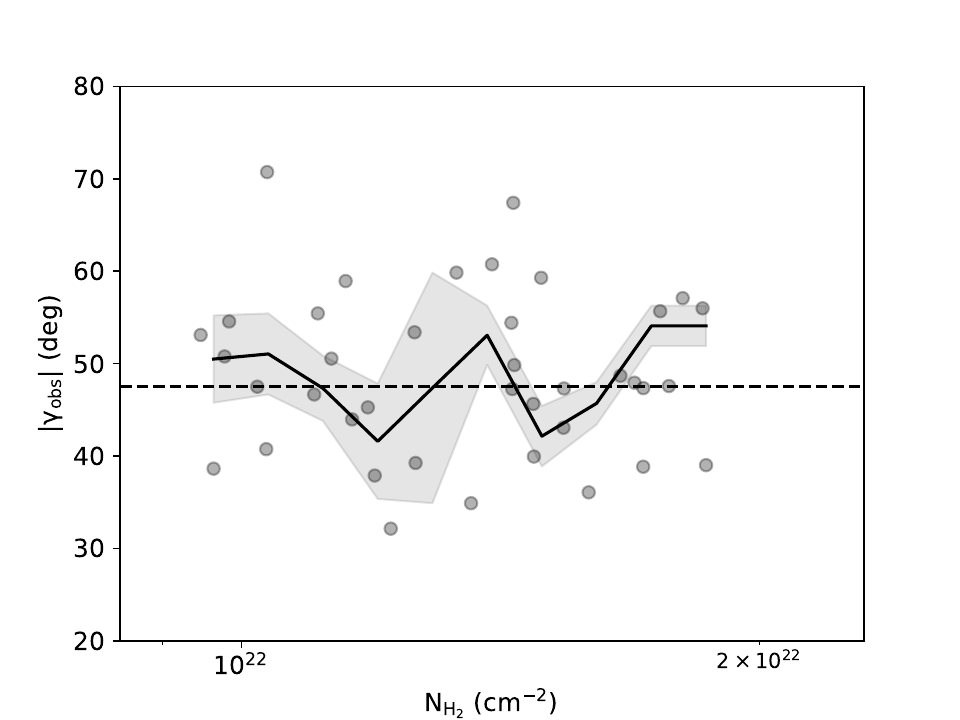}
    \caption{Left panel: Map of inferred inclination angles derived from the starlight polarization efficiency $P_{\rm K}/N_{\rm H}$ using the new technique of \cite{TruongHoang2025}. The segments demonstrate the mean polarization angles calculated in $2 ' \times 2 '$ grid cells, while the color scales represent the values of the inferred angles $|\gamma_{\rm obs}|$ within these grids. Right panel: The variation of inferred inclination angles $|\gamma_{\rm obs}|$ with respect to the column density $N_{\rm H_2}$. Solid black line present the running mean of the data with $1\sigma$ uncertainty (dark shaded region), whereas the global mean of $|\gamma_{\rm obs}| \sim 48^{\circ}$ is demonstrated by a dashed black horizontal line.}
    \label{fig:Incl_Star}
\end{figure*}

Figure \ref{fig:Incl_Star} shows the spatial distribution of the absolute inclination angles $|\gamma_{\rm obs}|$ obtained from our technique (left panel). The right panel of Figure \ref{fig:Incl_Star} demonstrates the variation of $|\gamma_{\rm obs}|$ with respect to the column density $N_{\rm H_2}$. The global B-fields are inclined with respect to the LOS with the mean $|\gamma_{\rm obs}| \sim 48$ degrees, and not significant changed as $N_{\rm H_2}$ increases. The local values of $|\gamma_{\rm obs}|$, on the other hand, fluctuate strongly owing to the local variations of 3D B-field structures from the outer to the inner regions of G11.

\subsection{Constructing the 3D B-field Morphology}
\label{sec:result_3Dstruc}
The main advantage of inferring 3D inclination angles from dust polarization degree is that we can reconstruct the 3D B-field morphology in star-forming regions when combining with the 2D POS B-field patterns observed in the dust polarization map. Here, we present the possible 3D B-field structures in the outer regions of G11 derived from the inferred inclination angles.

\begin{figure*}
    \includegraphics[width=0.35\linewidth]{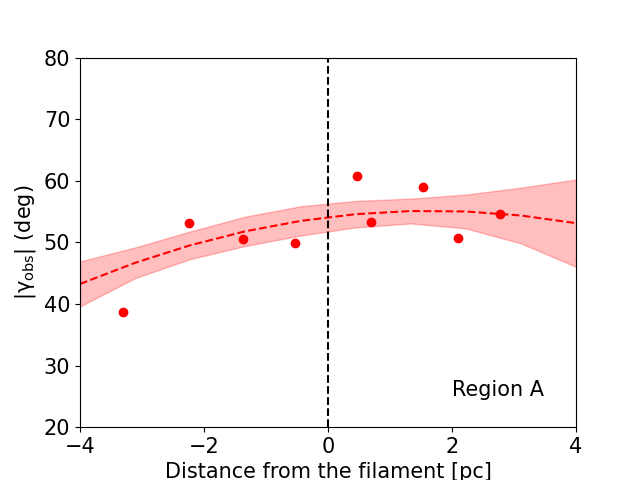}
    \includegraphics[width=0.35\linewidth]{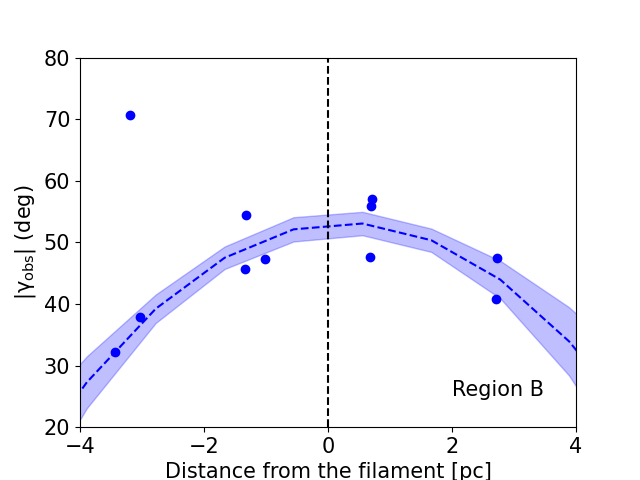} 
    \includegraphics[width=0.35\linewidth]{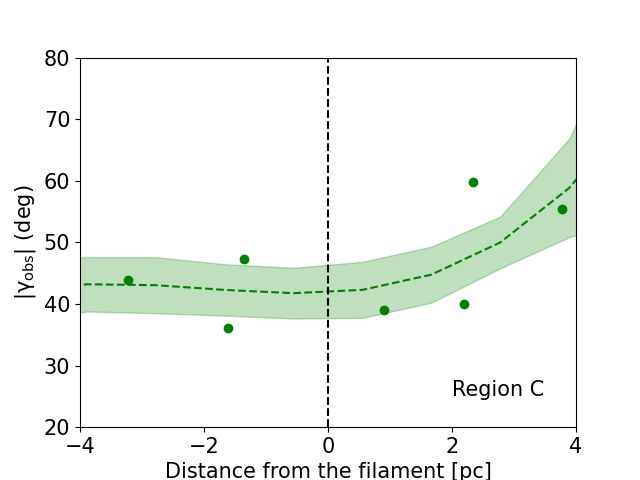}
    \caption{The variation of inferred inclination angles $|\gamma_{\rm obs}|$ with respect to the distance from the filament spine for Regions A (left panel), B (middle panel), and C (right panel). A polynomial fit is applied to fit the data with $1\sigma$ uncertainty (color shade regions).}
    \label{fig:incl_filament}
\end{figure*}

\begin{figure*}
    \includegraphics[width=1.0\linewidth]{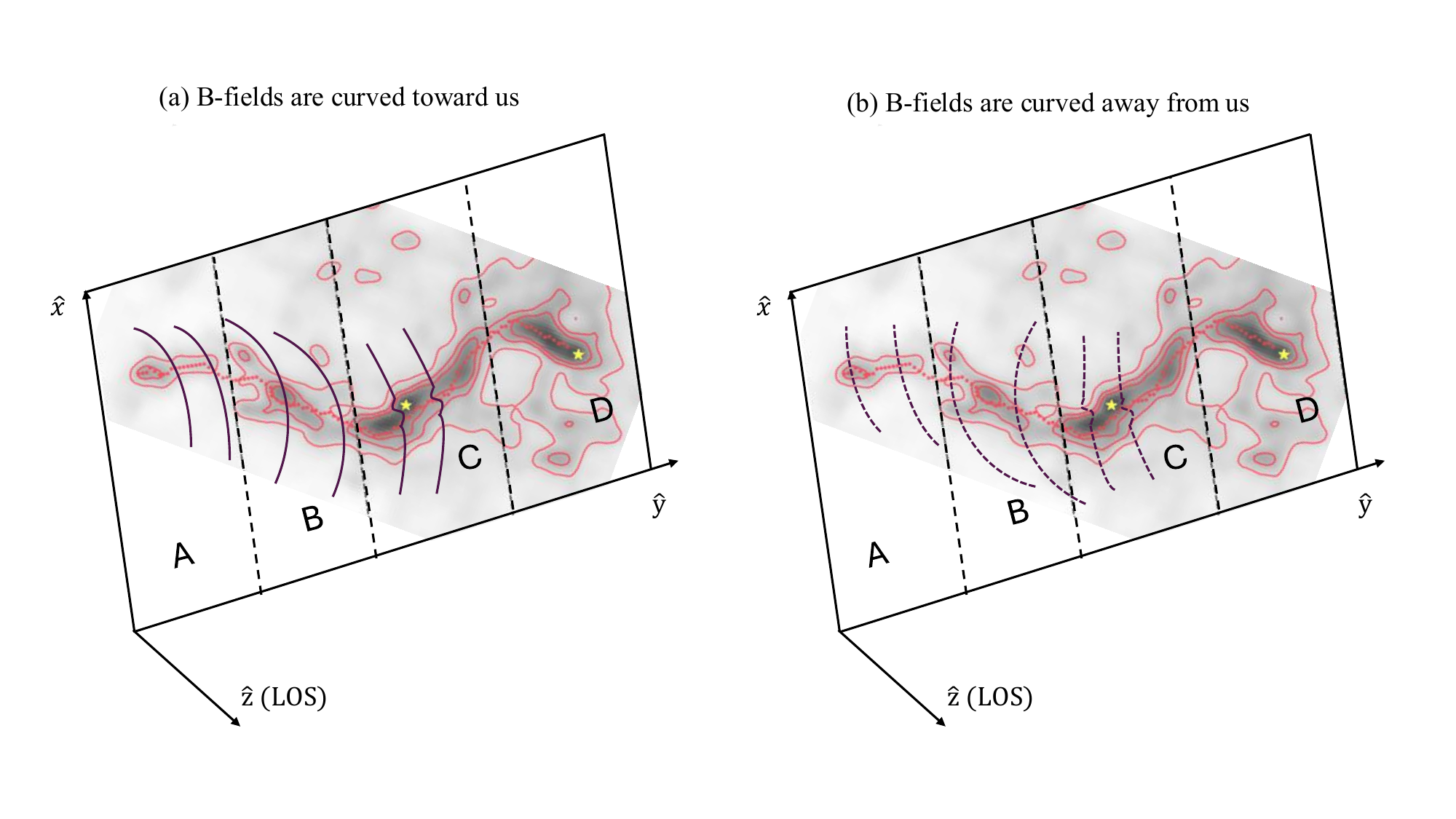}
    \caption{Illustration of the morphology of 3D B-fields in Regions A, B and C of the G11 filament derived from the inferred inclination angles in two scenarios: (a) 3D B-fields are curved toward the observer with the positive $\gamma_{\rm obs}$ and (b) 3D B-fields are curved away from the observer with the negative $-\gamma_{\rm obs}$.}
    \label{fig:3D_G11}
\end{figure*}

Figure \ref{fig:incl_filament} shows the variation of the absolute inclination angles $|\gamma_{\rm obs}|$ across the filament spine in Regions A (left panel), B (middle panel), and C (right panel). The POS magnetic fields are roughly perpendicular to the filament structure in three regions, as observed from both starlight polarization observations (\cite{Chen2023}, see Figure \ref{fig:Pol_Ks}) and thermal dust polarization observations by \cite{Ngoc2023}. A polynomial fit was applied to the data points with $1\sigma$ uncertainties (color-shaded regions). The illustration of the possible 3D B-fields morphologies derived from the inferred angles is demonstrated in Figure \ref{fig:3D_G11}.  

One can see that, in Region A where the gas column density is low ($n_{\rm H_2} \sim 600 - 700\,\cm^{-3}$, see Figure \ref{fig:nH2_Td}), the mean inclination angle is around 45 degrees in the outer regions, and tends to increase to 50 degrees in the dense filament spine. The profile of $|\gamma_{\rm obs}|$ is similar in Region B but with a larger deviation across the filament (e.g., $|\gamma_{\rm obs}| \sim 30^{\circ}$ in the outer regions and $|\gamma_{\rm obs}| \sim 50^{\circ}$ in the inner dense filament). From these profiles, we suggest that there is evidence of an arc-shaped B-field morphology with different curvature that wraps the G11 filament. Such arc-shaped 3D B-fields are also observed in other large-scale filamentary clouds such as Perseus and Orion A by \cite{Tahani2018, Tahani2022} or in small-scale star-forming regions as Orion Molecular Cloud 1 (OMC-1) by \cite{Tram2024}. The derived inclination angles are absolute values; therefore, there could be two scenarios of 3D B-field orientation: (1) the arc-shaped fields are curved toward the observer with the positive $\gamma_{\rm obs}$ (left panel) and (2) away from the observer with the negative $-\gamma_{\rm obs}$ (right panel).

In Region C, on the other hand, the results of absolute inclination angles show an opposite trend, with a decrease in $|\gamma_{\rm obs}|$ from 50 degrees to $\sim 40$ degrees toward the filament spine. The decrease in $|\gamma_{\rm obs}|$ could be caused by the back-and-forth bending of 3D B-fields due to the strong effect of gravitational contraction - same as the results from the synthetic polarization data of filamentary clouds undergoing gravitational collapse by \cite{TruongHoang2025}. 

We note that in Region D, the magnetic fields are shown to be parallel to the filament structure (\citealt{Chen2023}). These POS patterns differ from the perpendicular patterns of B-fields found in the inner dense filament by SOFIA/HAWC+ by \cite{Ngoc2023}. The differences in the POS B-field morphology could result from tracing different dust populations between the outer regions and the inner filament spine, raising the complexity when constraining 3D B-fields by starlight and thermal dust polarization. In the upcoming study, we will apply our new technique to multi-wavelength thermal dust polarization and investigate the 3D B-field morphology toward Region D in more detail (Ngoc et al., in preparation).

\section{Discussion}
\label{sec:discuss}
\subsection{3D B-field Strength and Energy Balance}
\label{sec:discuss_3Dstrength}
Typically, the strength of $B_{\rm POS}$ can be derived from dust polarization observations by taking the distortion of local B-fields perpendicular to the mean fields induced by gaseous turbulence (i.e., the turbulent magnetic energy is equal to the turbulent gas kinetic energy). By introducing the inferred B-field inclination angles with respect to the LOS from the polarization degree, $\gamma_{\rm obs}$, the 3D B-field strength can be inferred from dust polarization observations itself as (\citealt{HoangTruong2024})
\begin{equation}
    B_{\rm 3D} = \frac{B_{\rm POS}}{\sin\gamma_{\rm obs}},
\label{eq:B_3d}
\end{equation}
where the magnitude of $B_{\rm POS}$ can be estimated from the dispersion of POS polarization angle, $\sigma_{\theta}$, using the original Davis-Chandrasekhar-Fermi (DCF) formulation (\citealt{Davis1951}; \citealt{Chandrasekhar1953}) as follows
\begin{equation}
    B_{\rm POS, DCF} = Q\sqrt{4\pi\rho}\frac{\sigma_{v}}{\sigma_{\theta}},
\label{eq:B_pos}
\end{equation}
where $Q = 0.5$ is the correction factor following by \cite{Ostriker2001} for $\sigma_{\theta} < 25^{\circ}$, $\rho = \mu m_{\rm H} n_{\rm H_2}$ is the mass density of molecular hydrogen with $\mu = 2.83$ (\citealt{Kauffmann2008}) and $\sigma_{v}$ is the non-thermal gas velocity dispersion. 

The original DCF formulation is based on the assumption of incompressible turbulence. There is also the presence of compressive turbulence in the diffuse ISM and star-forming regions (\citealt{Heiles2003}; \citealt{ChoLaz2003}; \citealt{SkaliddisTassis2021}; \citealt{Skalidis2021}). \cite{SkaliddisTassis2021} took into account the compressible mode of ISM turbulence in the estimation of POS B-field strength (hereafter ST method), which is given by
\begin{equation}
    B_{\rm POS, ST} = \sqrt{2\pi\rho}\frac{\sigma_{v}}{\sqrt{\sigma_{\theta}}}.
\label{eq:B_pos_ST}
\end{equation}

We adopt the polarization angle dispersion for each grid cell spacing of $2' \times 2'$ (Section \ref{sec:Fturb}) and the volume density map of molecular hydrogen $n_{\rm H_2}$ in the outer regions of G11 (Section \ref{sec:nH2_Td}). We adopt the velocity dispersion from the available moment 2 map of $\rm ^{13}CO$ J=1-0 emission of G11 in the velocity range of $25 - 37\rm\, km\,s^{-1}$, which was obtained from the Milky Way Imaging Scroll Painting (MWISP) project - a northern Galactic plane CO survey using the 13.7 m telescope of the Purple Mountain Observatory (see \citealt{Su2019}). The thermal contribution was removed, so that the non-thermal dispersion by turbulence is $\sigma_v = \sqrt{\delta v_{\rm obs}^2 - c_s^2}$, where $c_s \approx 0.2\,\rm km\,s^{-1}$ for the constant gas temperature of $T_{\rm ex} \approx 12\,\rm K$ derived from the  $\rm ^{12}CO$ emission toward G11 (\citealt{Chen2023}). The mean non-thermal velocity dispersion in each cell of $2' \times 2'$ size was calculated from $\rm ^{13}CO$ data by \cite{Chen2023} (see also their Table 2). The B-field strength on the POS is then calculated from the DCF (Equation \ref{eq:B_pos}) and ST methods (Equation \ref{eq:B_pos_ST}). Given the local B-field inclination angles inferred from the observed starlight polarization efficiency (Section \ref{sec:inferred_incl}), we obtain the local 3D B-field strength $B_{\rm 3D}$ as shown in Equation \ref{eq:B_3d}.

\begin{figure}
    \includegraphics[width=1.0\linewidth]{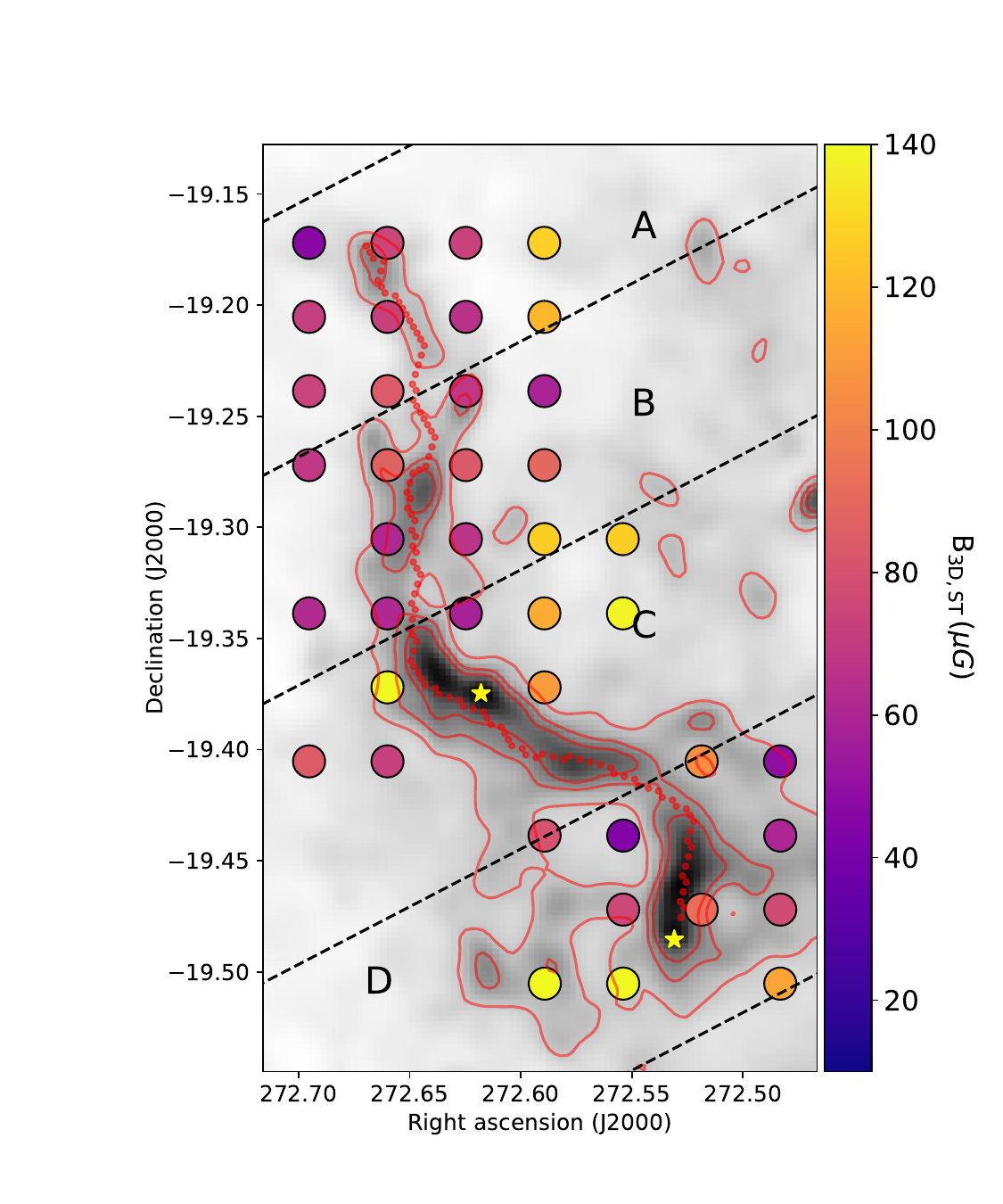}
    \caption{The spatial distribution of the local 3D B-field strength $B_{\rm 3D}$ in each $2' \times 2'$ grid cell in the G11 filament. The POS B-fields are calculated from the ST method considering compressive turbulence effect (\citealt{SkaliddisTassis2021}).}
    \label{fig:B3D}
\end{figure}

Figure \ref{fig:B3D} shows the results of the local 3D B-field strength $B_{\rm 3D}$ in the G11 filament. The $B_{\rm POS}$ from the ST method and the inclination angle effect are taken into consideration. Table \ref{tab:B_strength} summarizes the mean strength of $B_{\rm POS}$ and $B_{\rm 3D}$ in each region of G11. The mean B-field strength calculated from the ST method are presented within parentheses. The mean POS B-field strength from the original DCF method in the outer regions of G11 is around $60 - 100\,\rm\mu G$. Once the ST method is applied, the result is lower and nearly uniform in four regions due to the weakening effect of $\sigma_{\theta}$ variation ($B_{\rm POS} \sim 52 - 70\,\rm\mu G$). By taking the inclination angle effect, the 3D B-field strength is higher by a factor of $\sim 1.25 - 1.35$ for both results from two techniques. The derived factors are different than the statistical derivation of $B_{\rm 3D} \approx 1.27\,B_{\rm POS}$ by \cite{Crutcher2004} for a magnetic field geometry without a preferred inclined angle. This implies the importance of 3D inclination angles inferred from dust polarization in improving 3D B-field strength estimation in star-forming regions. 

\begin{table*}[]
    \centering
    \caption{Summary of the mean POS B-field strength $\langle B_{\rm POS}\rangle$, the 3D strength $\langle B_{\rm 3D} \rangle$ and the derived mean \alfvenic Mach number $\langle M_A \rangle$ and the critical mass-to-flux ratio $\langle \mu_\phi \rangle$ in four sub-regions in the G11 filament.}
    \begin{tabular}{l c c c c c c c c }
    \toprule
      Region & $\langle n_{\rm H_2} \rangle$  & $\langle \sigma_v \rangle$  & $\langle \bar{\sigma}_\theta \rangle$ &$\langle |\gamma_{\rm obs}| \rangle$& $\langle B_{\rm POS} \rangle$ & $\langle B_{\rm 3D}\rangle$  & $\langle M_{A} \rangle$ & $\langle \mu_{\phi} \rangle$ \\

      & ($\rm cm^{-3}$) & ($\rm km\,\,s^{-1}$) & (deg) & (deg)&($\mu\rm G$) &($\mu\rm G$) & &  \\

      \midrule

      A & 722.7 & 1.75 & 10.7 & 51.9 & 95.6 (59.5)  & 119.5 (73.6) & 0.3 (0.4) & 1.12 (1.82) \\

      B & 860.8 & 1.8 & 15.1 & 47.5 & 73.3 (52.8)  & 98.4 (68.3) & 0.36 (0.55) & 1.8 (2.41)\\

      C & 864.2 & 2.0 & 13.4 & 44.0 & 108.3 (69.5)  & 142.8 (109.6) &0.37 (0.49) & 1.03 (1.36)  \\

      D & 1086.4 & 2.05 & 28.0 & 47.6 & 63.0 (63.7)  & 84.5 (84.3) & 0.7 (0.7) & 2.7 (2.8) \\
  
      \bottomrule
    \end{tabular}
    \label{tab:B_strength}
\end{table*}

We re-investigate the interplay between the magnetic field, turbulence, and gravity in the G11 filament when the 3D local inclination angles are determined. We first quantify the \alfvenic Mach number, denoted by $M_{A}$, that characterizes the magnetic energy relative to the kinetic turbulent energy. The local $M_{A}$ of the gas is calculated as
\begin{equation}
    M_A = \frac{\sigma_v}{v_A},
\end{equation}
where $v_A = B_{\rm 3D}/\sqrt{4\pi\rho} = B_{\rm POS}/(\sin{\gamma_{\rm_{obs}}}\sqrt{4\pi\rho})$ is the \alfvenic velocity. $M_{\rm A} < 1 $ indicates that the environment is magnetic-dominated (i.e., sub-\alfvenic  turbulence), whereas $M_{\rm A} > 1 $ indicates that the environment is turbulence-dominated (i.e., super-\alfvenic  turbulence). The $M_A$ depends only on the polarization angle dispersion and the 3D inclination angles as $M_{A} \varpropto \sin{\gamma_{\rm obs}}\sigma_{\theta}$ (for $B_{\rm POS}$ from the DCF method with a fixed $Q$, see Equation \ref{eq:B_pos}), or $M_{A} \varpropto \sin{\gamma_{\rm obs}}\sqrt{2\sigma_{\theta}}$ (for $B_{\rm POS}$ from the ST method, see Equation \ref{eq:B_pos_ST}).

We then quantify the mass-to-flux ratio, denoted by $M/\Phi$, which describes how 3D B fields could regulate star formation processes against the effect of self-gravity (\citealt{Nakano1978}). The calculation of mass-to-flux ratio in the unit of critical value is given by (\citealt{Crutcher2004_Aps})
\begin{equation}
    \mu_{\phi} = \frac{(M/\Phi)_{\rm obs}}{(M/\Phi)_{\rm crit}} = 7.6 \times 10^{-21}\frac{N_{\rm H_2, flux}}{B_{\rm flux}}, 
\end{equation}
where $N_{\rm H_2, flux}$ and $B_{\rm flux}$ are the column density of molecular hydrogen gas and the total B-field strength along a single flux tube. We assume the flux tube is parallel to the LOS, so that $N_{\rm H_2, flux} = N_{\rm H_2}$ obtained from the Herschel observations (Section \ref{sec:nH2_Td}), and $B_{\rm flux} = B_{\rm LOS} = B_{\rm POS}/\tan{\gamma_{\rm obs}}$. This is the upper limit of the 3D mass-to-flux ratio that can be achieved from the LOS field strength (\citealt{Tritsis2026}). Small $\mu_{\phi} < 1$ indicates that B-fields support the filament structure against the gravitational collapse (i.e., sub-critical), while large $\mu_{\phi} > 1$ demonstrates that the B-fields cannot prevent the gravitational collapse to form a new star (i.e., super-critical). Here, we re-calculate both values of $M_{\rm A}$ and $\mu_{\rm \phi}$ in each $2' \times 2'$ cell when the local 3D inclination angles $\gamma_{\rm obs}$ inferred from dust polarization are considered.

\begin{figure*}
    \includegraphics[width=0.5\linewidth]{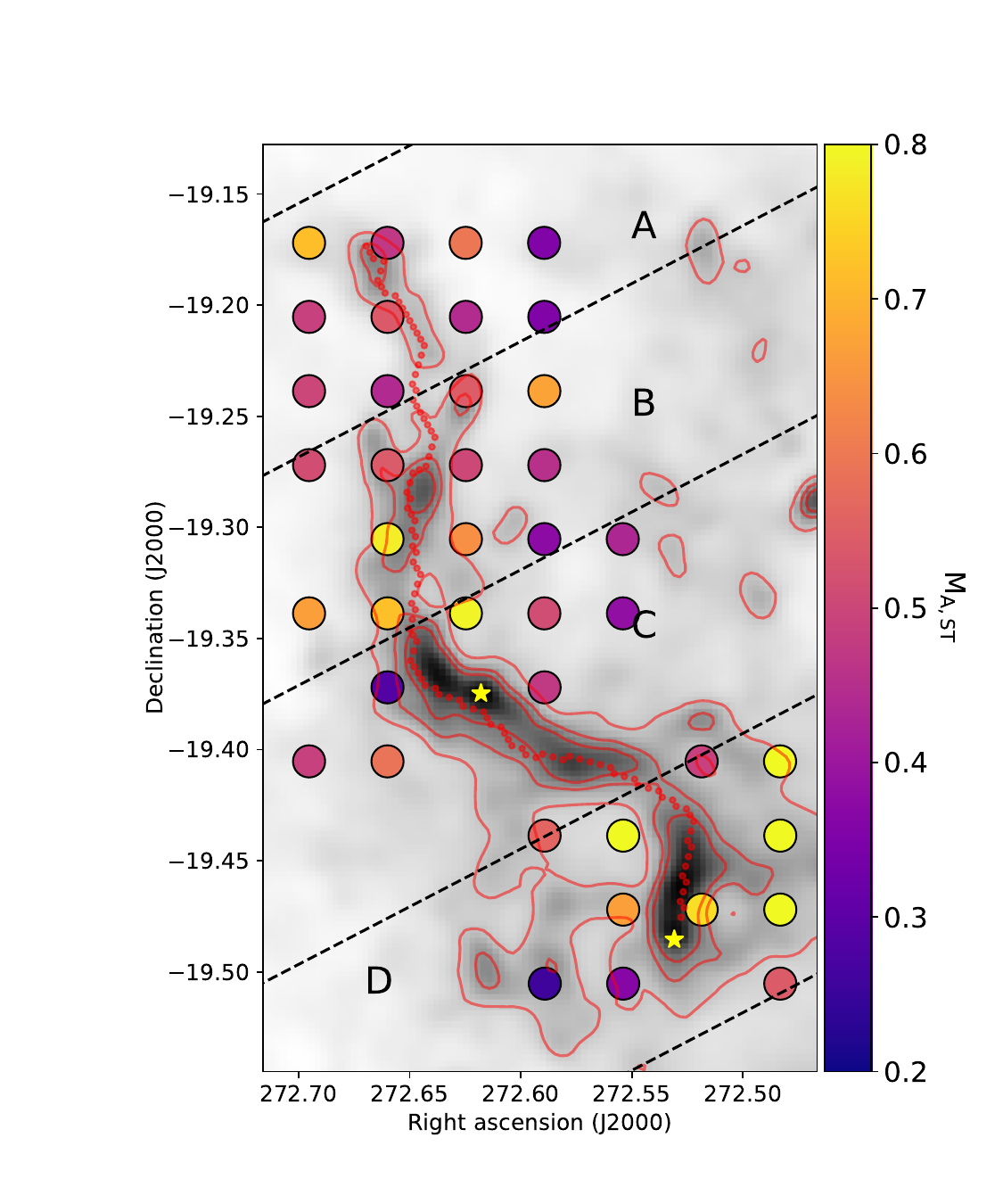}
    \includegraphics[width=0.5\linewidth]{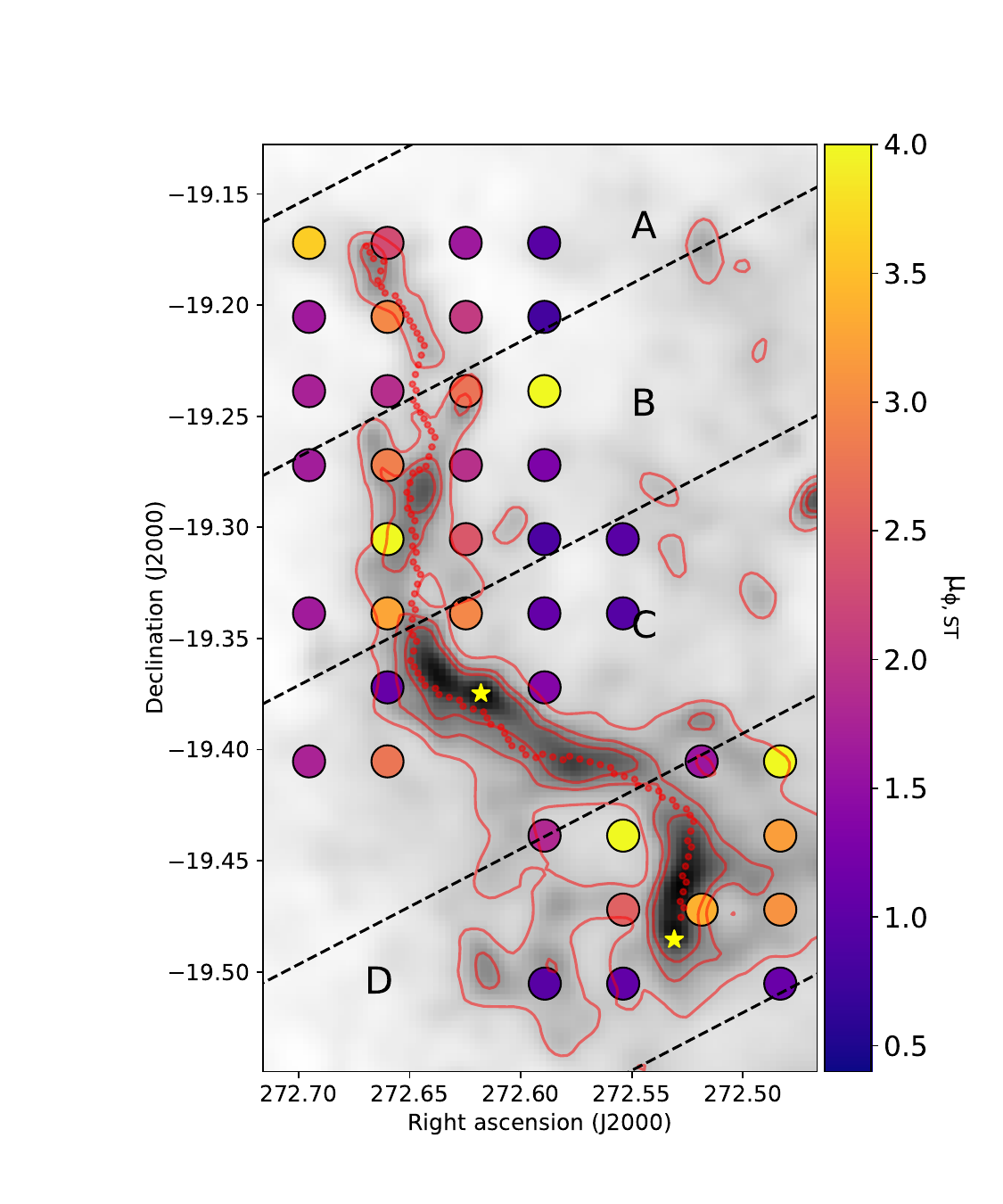}
    \caption{The maps of the \alfvenic Mach $M_{\rm A}$ (left panel) and the critical mass-to-flux ratio $\mu_{\rm \phi}$ (right panel) when the $B_{\rm POS}$ from the ST method (\citealt{SkaliddisTassis2021}) and the effect of 3D inclination angles are considered.}
    \label{fig:mu_MA}
\end{figure*}

Figure \ref{fig:mu_MA} shows the maps of $M_{\rm A}$ (left panel) and $\mu_{\rm \phi}$ (right panel) within G11 when the $B_{\rm POS}$ is calculated from the ST method and the inclination angle effect is taken into consideration. The mean $\langle M_{A} \rangle$ and $\langle \mu_{\phi}\rangle$ in each region of G11 derived from the POS B-field strength from DCF and ST methods (within parentheses) and the local inclination angles are also shown in Table \ref{tab:B_strength}. The variation of $M_{\rm A}$  and $\mu_{\phi}$ are consistent as \cite{Chen2023}: the outer regions of G11 is mostly sub-\alfvenic, sub- and trans-critical, with low $M_{\rm A} < 1$ and $\mu_{\rm \phi} \lesssim 1$, and becomes super-critical in high-density filaments (i.e., $\mu_{\rm \phi} > 2$) where the gravitational contraction is dominated to form massive clumps and dense cores (\citealt{Henning2010}; \citealt{Kainulainen2013}; \citealt{Wang2014}). However, as the effect of local inclination angles is considered, the dynamical importance of B-fields in local environments is more emphasized. This results in lower $M_{\rm A}$ and $\mu_{\rm \phi}$, in comparison to the derived values by \cite{Chen2023}. 

\subsection{Implications of 3D B-fields Morphology for Understanding Filament Evolution}
From the local 3D inclination angles inferred from the starlight polarization efficiency (Section \ref{sec:result_3Dstruc}), we can derive the 3D B-field structures as shown in Figure \ref{fig:incl_filament} for three different regions, A, B, and C, where the POS B-fields are shown to be perpendicular to the filament spine (\citealt{Chen2023}). Here, we discuss in detail the variation of 3D B-field morphology and its implications for understanding filament evolution.

In Regions A and B, we found the increase in the inferred angles toward the dense filament spine, suggesting the 3D arc-shaped B-field morphology (see Figure \ref{fig:3D_G11}). However, the 3D curvature of the arc-shaped structure could be different between the two regions. For Region A, the difference in the inferred angles between the outer regions and the inner dense filament is small, and the 3D B-fields are less curved toward the LOS and nearly parallel to the POS. The mean fields, therefore, could regulate gas flows from the envelope to the spine and less produce velocity gradient along the sightline, as previously found in the moment 1 map from the $\rm ^{13}CO$ J=1-0 emission by \cite{Chen2023}. For Region B, the deviation between the outer and inner regions is significant, suggesting the arc-shaped B-field structure with high curvature toward the LOS. This could be associated with the red-shifted and blue-shifted LOS components of gas flows threaded by the highly curved B-fields observed from the $\rm ^{13}CO$ J=1-0 emission data (see \citealt{Chen2023}). The arc-shaped B-fields could point toward or away from the observer depending on the sign of the derived angles $|\gamma_{\rm obs}|$ (left and right panels of Figure \ref{fig:3D_G11}). The exact orientation of the 3D arc-shaped B-fields could be determined by combining the inferred angles from dust polarization with the LOS directions of B-fields provided by Faraday rotation (see, e.g., \citealt{Tahani2018, Tahani2019, Tahani2022}) or by Zeeman splitting of spectral lines (see \citealt{Reissl2018}).

The finding of the 3D arc-shaped B-field morphology from the inferred angles is crucial for understanding the formation and evolution of filamentary clouds in the magnetized medium. The origin of 3D arc-shaped structure could be a result of the interaction between the cloud and supernova shock that bends initial field lines and forms a filamentary structure (\citealt{Inutsuka2015}; \citealt{Inoue2018}; \citealt{Tahani2018, Tahani2019}). However, there is no evidence of external feedback (e.g., supernova feedback) located behind G11, providing uncertainty in confirming this scenario. Another possibility is the presence of gaseous accretion flows along the LOS driven by gravity that causes the B-field dragging in high-density filamentary clouds (e.g., U-shape structure, see \citealt{Motte2018}; \citealt{Gomez2018}; \citealt{Pillai2020}; \citealt{Tapinassi2024}), which could be favored as the impact of gravity becomes dominant in Region B with large $\mu_{\phi} > 2$ (see Figure \ref{fig:mu_MA}).

In Region C, the gravitational contraction is more prominent, resulting in the formation of the massive clump P1 and multiple cores (\citealt{Henning2010}; \citealt{Kainulainen2013}; \citealt{Wang2014}; \citealt{Dewangan2024}). The strong gravitational effect could alter the 3D B-field morphology, causing the back-and-forth bending of B-fields with respect to the sightline and the decrease in the inferred angles to the filament spine as shown in Figures \ref{fig:incl_filament} and \ref{fig:3D_G11}). In future studies, we will combine both the multi-scale 3D B-field observations from dust polarization and the kinematics information from different gas tracers to fully explore the origins of 3D B-field structure and their impacts on massive filament formation and evolution.

\subsection{Potential Effect of Anisotropic Grain Growth}
\label{sec:discuss_graingrowth}
We emphasize that the constraint on intrinsic dust properties (e.g., grain size distribution and elongation) is strongly associated with the accurate inference of 3D inclination angles from starlight and thermal dust polarization degree (\citealt{HoangTruong2024}; \citealt{TruongHoang2025}). We utilized the observations of optical dust extinction provided by the Gaia mission (\citealt{Zhang2025}) to derive the constrained grain size distribution mostly in the outermost region of G11. Using the observed averaged optical $\langle R_{\rm V}\rangle \sim 2.83$, we derive the mean maximum grain size $\langle a_{\rm max}\rangle = 0.25\,\rm\mu m$. This is the lowest limit of $a_{\rm max}$ that can be presented in the outermost regions of G11 with $N_{\rm H_2} \lesssim 10^{22}\,\rm cm^{-2}$ (Section \ref{sec:int_amax}). Combined with the maximum starlight polarization efficiency achieved in \cite{Chen2023}, we then derived the lower limit of grain axial ratio of $s \gtrsim  1.4$ (Section \ref{sec:int_elongation}). Our derived dust properties are similar to those found in the diffuse ISM observations (\citealt{Mathis1977}; \citealt{Hensley2023}). These dust properties are considered to be uniform in the outer regions of G11 and make a significant contribution to the derivation of 3D inferred angles $\gamma_{\rm obs}$ from starlight polarization efficiency (Section \ref{sec:inferred_incl}). 

Nevertheless, it is suggested that these dust properties can vary with respect to local gas density as a result of anisotropic grain growth: (1) increasing $a_{\rm max}$ (\citealt{Hirashita2012}; \citealt{Bate2022}) and (2) increasing grain elongation $s$ (\citealt{Hoang2022_Graingrowth}). In the inner dense filament with the filament width of $1\,\rm pc$ and $n_{\rm H_2}$ in the magnitude orders of $\sim 10^4\,\cm^{-3}$, \cite{Ngoc2023} found the increased alignment size to $a_{\rm align} \sim 0.3\,\rm\mu m > a_{\rm max} $ requiring for reproducing the current thermal dust polarization degree by SOFIA/HAWC+ at $214\,\rm\mu m$. This indicates the signature of moderate growth in grain size to $a > 0.3\,\rm\mu m$ presented in the densest part of the G11 filament. The grain axial ratio is expected to increase to $s \gtrsim 2$ to explain the thermal dust polarization data (Ngoc et al, in preparation). The anisotropic grain growth could also be found in small-scale starless cores by \cite{Tram2025} with $a_{\rm max}$ up to $\sim 1 - 1.5\,\rm\mu m$ and $s$ up to 3. 

The anisotropic grain growth has a significant impact on the determination of dust properties and 3D B-fields from both dust extinction and polarization. The increasing $a_{\rm max}$ could possibly cause higher extinction at longer wavelengths and higher $R_{\rm V}$ (see Appendix \ref{sec:appendix_extinction}). This could explain the high optical $R_{\rm V} \approx 3.5$ found in the outermost of Regions C and D with $n_{\rm H_2} \sim 1-2 \times 10^3\,\rm cm^{-3}$ (Figure \ref{fig:Rv_obs}), corresponding to $a_{\rm max} \sim 0.3\,\rm\mu m$ (Figure \ref{fig:Rv}). The $R_{\rm V}$ is expected to increase more dramatically in the inner dense filament with $N_{\rm H_2} > 10^{22}\,\rm cm^{-2}$ ($n_{\rm H_2} > 10^3\,\rm cm^{-3}$); however, the detection of dust extinction from background stars at NIR and MIR wavelengths is required for further evaluation (\citealt{Butler2009}; \citealt{Kainulainen2013}; \citealt{Juvela2026}). Additionally, the change in both $a_{\rm max}$ and grain axial ratio $s$ could provide degeneracy when calculating inferred inclination angles from available dust polarization data (\citealt{HoangTruong2024}, see Appendix \ref{sec:appendix_degeneracy} for quantitative calculations). We will provide the better constraint on dust properties in the inner dense filament of G11 by combining the available polarization observations with the NIR/MIR dust extinction from available data from Spitzer (\citealt{Butler2009}; \citealt{Kainulainen2013}) or future observations from James Webb Telescope (JWST, see \citealt{Wang2024}; \citealt{Ferres2025}); or the dust spectral index from FIR/sub-mm thermal dust emission (\citealt{Draine2006}; \citealt{Juvela2015}). These can provide a better inference of 3D inclination angles on multiple scales of the G11 filament and reach our ultimate goal of studying 3D B-field strength and morphology, and dust evolution in a follow-up study (Ngoc et al., in preparation).

\subsection{Effect of Iron Inclusions}
\label{sec:discuss_iron}
We have demonstrated the constraints on dust properties and 3D inclination angles of the mean B-fields in G11 from starlight polarization observations in the case of efficient magnetic alignment of large grains $a > a_{\rm align}$ with $R = 1$ (i.e., Ideal RAT alignment). In reality, the magnetic alignment by RATs relies on the magnetic properties of dust grains. \cite{Hoang2016a} demonstrated that for grains with paramagnetic (PM) materials (i.e., iron is randomly distributed inside grains), only RATs itself could not bring them to efficient alignment with respect to B-fields. The presence of iron inclusions embedded inside superpamagnetic (SPM) grains could enhance the efficiency of magnetic alignment via the combined effects of magnetic relaxation and RAT alignment (i.e., MRAT alignment, see \citealt{Lazarian2008}; \citealt{Hoang2016a}). Synthetic dust polarization modeling by \cite{Giang2023,Giang2024} showed the importance of MRAT alignment induced by SPM grains with iron inclusions to understand the high degree of polarization in protostellar environments with $n_{\rm gas} > 10^6\,\rm cm^{-3}$ observed by ALMA.

The analysis of \cite{TruongHoang2025} demonstrated the increase in $R$ when the number of iron atoms per cluster $N_{\rm cl}$ increases due to the enhancement of MRAT alignment for large SPM grains $a > a_{\rm align}$. Here, we calculate the Rayleigh reduction factor $R$ over the grain size for different values of $N_{\rm cl}$ (\citealt{Hoang2016a}; \citealt{TruongHoang2025}, see Equations \ref{eq:Ralign} - \ref{eq:delta_m} in Appendix \ref{sec:appendix_RAT}) and the alignment function $f_{\rm align}(a)$ as shown in Equation \ref{eq:falign}, using the \textsc{DustPOL\_py} code incorporating the modern MRAT theory (\citealt{Tram2021,Tram2024}). The environmental properties of G11 are derived from multi-wavelength Herschel observations (i.e., $n_{\rm H_2}$ and $T_{\rm d}$, see Figure \ref{fig:nH2_Td}). The constant $\langle B_{\rm 3D} \rangle \approx 80\,\rm\mu G$ is considered in the modeling. This assumption is reasonable for the uniform B-field strength found in the outer regions of G11 (see Section \ref{sec:discuss_3Dstrength}). The varying number of iron per cluster, $N_{\rm cl}$, and the volume filling factor $\phi_{\rm sp} = 0.01$ are assumed (i.e., corresponding to $3\%$ of iron incorporated into dust as a form of iron inclusions, see \citealt{Hoang2016a}). Following the procedure in Section \ref{sec:int_elongation}, we re-perform the numerical modeling of starlight polarization efficiency as shown in Equation \ref{eq:Pmod} with the $f_{\rm align}(a)$ derived from the MRAT theory. Fixed $\langle a_{\rm max} \rangle = 0.25\,\rm\mu m$ from the optical dust extinction (Section \ref{sec:int_amax}) and the ideal conditions of B-fields (i.e., $\sin^2\gamma = 1$ and $F_{\rm turb} = 1$) are being assumed at the maximum point. We compare with the observed maximum value (Figure \ref{fig:PK_max}) to constrain both the grain axial ratio and the number of iron inclusions for SPM grains presented in G11.

\begin{figure*}
    \includegraphics[width=0.35\linewidth]{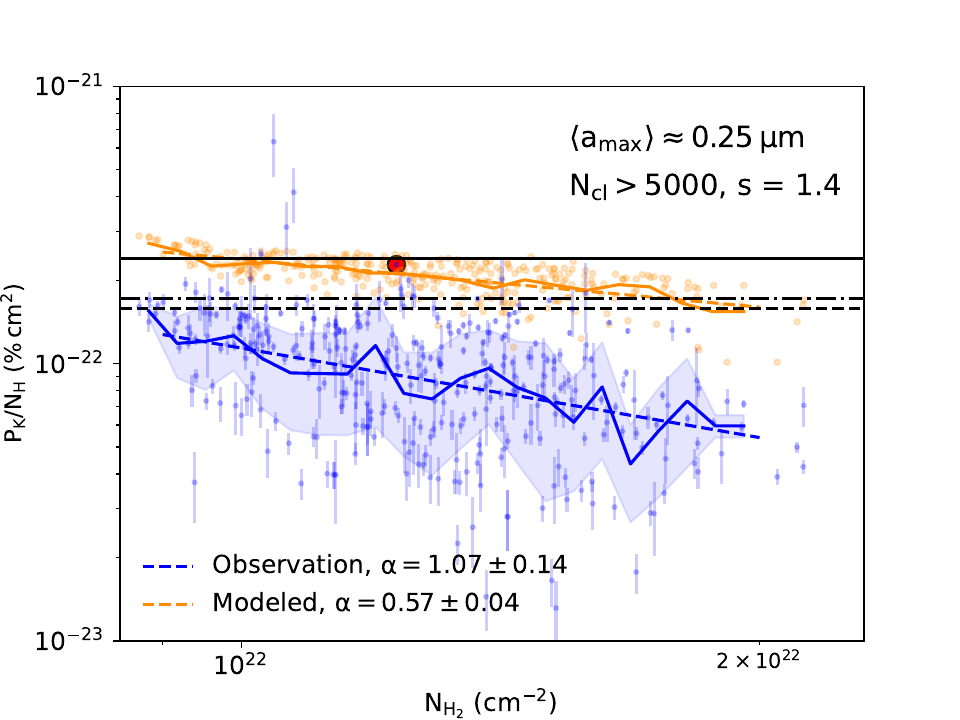}
    \includegraphics[width=0.35\linewidth]{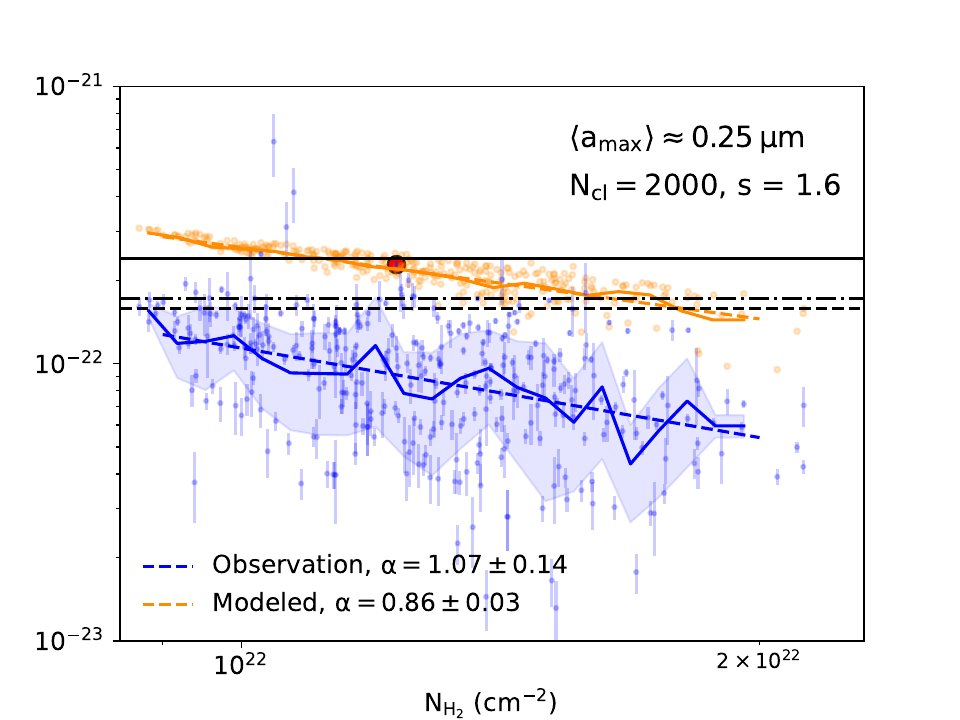} 
    \includegraphics[width=0.35\linewidth]{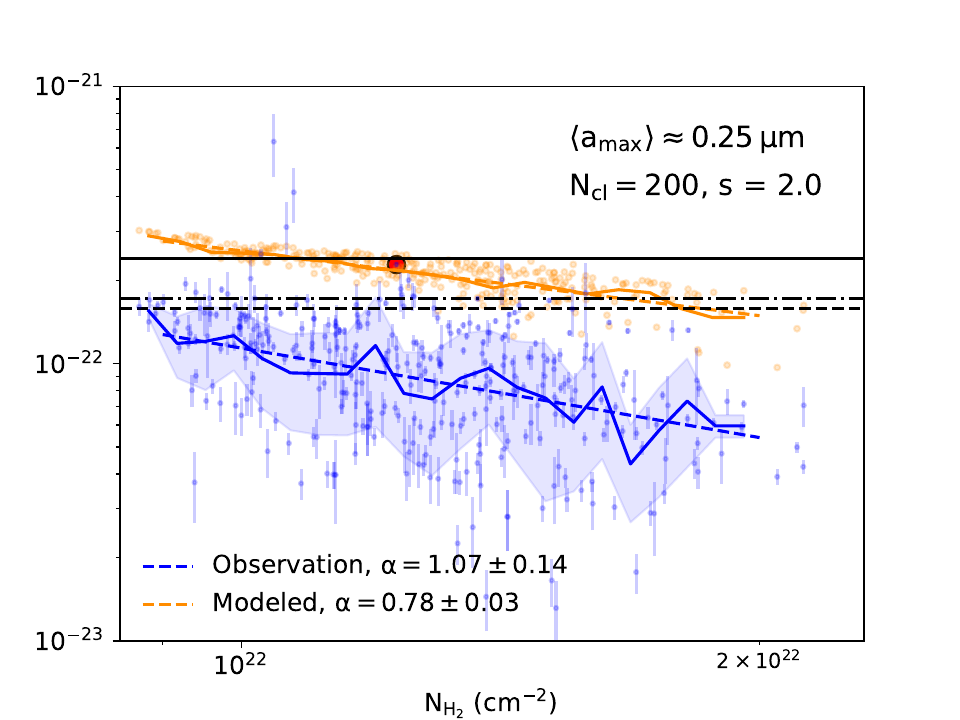}
    \caption{Observed polarization efficiency vs. the modeled one $(P_{\rm K}/N_{\rm H})_{\rm mod}$ generated by the \textsc{DustPOL\_py} code, but considering MRAT alignment for SPM grains with varying number of iron inclusions $N_{\rm cl}$ and grain elongations $s$. The ideal conditions of B-fields are considered at the maximum point, similar to those in Figure \ref{fig:PK_max}. The constrained values of $N_{\rm cl}$ and $s$ are determined when $(P_{\rm K}/N_{\rm H})_{\rm mod} \approx (P_{\rm K}/N_{\rm H})_{\rm max}$.}
    \label{fig:iron_constrain}
\end{figure*}

Figure \ref{fig:iron_constrain} demonstrates the constraints on magnetic properties of SPM grains and grain elongation by comparing the observed maximum value and modeled one when the MRAT alignment is taken into account. Table \ref{tab:Ncl_s_bestfit} summarizes the constrained values from the observed maximum polarization efficiency. The previous constraint on the grain axial ratio of $s \gtrsim 1.4$ can be derived from the maximum polarization efficiency if SPM grains achieve perfect magnetic alignment by MRAT with high levels of iron inclusion $N_{\rm cl} > 5000$. The MRAT alignment efficiency could be lower if grains have lower levels of iron inclusions. This effect is more prominent in the inner dense filament due to the reduction of magnetic relaxation by increasing gas randomization (see \citealt{Hoang2016a}), resulting in the steeper slope $\alpha >$ 0.5 when $N_{\rm cl}$ decreases to 200. The derived inclination angles $|\gamma_{\rm obs}|$ are then expected to increase in the inner dense filament due to the degeneracy with the reduced MRAT alignment efficiency. Besides this, the increase in grain axial ratio to $s \gtrsim 1.6$ or $s \gtrsim 2$ is required to retrieve the maximum polarization efficiency observed by \cite{Chen2023}. This raises the importance of characterizing the magnetic properties of grains to accurately characterize the alignment properties and constrain both 3D B-fields and dust properties from dust polarization observations.

\begin{table*}[]
    \centering
    \caption{Summary of intrinsic dust properties constrained from the observed maximum polarization efficiency, $(P_{\rm K}/N_{\rm H})_{\rm max}$, for both Ideal RAT and MRAT alignment models for SPM grains with iron inclusions.}
    \begin{tabular}{l c c}
    \toprule
      Alignment Model  &  Grain elongation ($s$) & Magnetic properties of grains  \\
      \midrule
      Ideal RAT  &1.4  & $R=1$ \cr
      MRAT   & 1.4 & SPM, $N_{\rm cl} > 5000$ \cr
          & 1.6 &  SPM, $N_{\rm cl} = 2000$ \cr
          & 2.0 &  SPM, $N_{\rm cl}= 200$ \cr
      \bottomrule
    \end{tabular}
    \label{tab:Ncl_s_bestfit}
\end{table*}



\subsection{Analysis Caveats}
\label{sec:discuss_caveats}
An accurate determination of 3D inclination angles from starlight polarization efficiency can be achieved only when the intrinsic dust properties, the efficiency of grain alignment and depolarization by magnetic turbulence are well constrained. However, these parameters exhibit various uncertainties.

Firstly, we made two key assumptions: (I) the intrinsic dust properties (axial ratio $s$ and $a_{\rm max}$) are uniform with a constant intrinsic polarization efficiency $P_{\rm i,K}/N_{\rm H}$ across the cloud (Section \ref{sec:int_fpol}); and (II) Ideal RAT alignment and Ideal conditions of B-fields (i.e., $\sin^2\gamma \sim 1$ and $F_{\rm turb} \sim 1$ at the maximum observed polarization efficiency $(P_{\rm K}/N_{\rm H})_{\rm max}$ (Section \ref{sec:int_elongation}). For Assumption I, grain growth in high-density regions (\citealt{Hirashita2012}; \citealt{Bate2022}; \citealt{Hoang2022_Graingrowth} could increase in both $a_{\rm max}$ and $s$ across G11 (see the discussion in Section \ref{sec:discuss_graingrowth}). This effect can increase the intrinsic polarization efficiency $P_{\rm i,K}/N_{\rm H}$, causing the reduction of $|\gamma_{\rm obs}|$ (i.e., 3D B-fields are more inclined to the LOS) inferred from the observed polarization efficiency $P_{\rm K}/N_{\rm H}$ (see Appendix \ref{sec:appendix_degeneracy}). For Assumption II, it is unknown whether B-fields lie in the POS in the location of maximum polarization efficiency, and thus it can be affected by the viewing angle between the mean B-fields and the LOS (see \citealt{TruongHoang2025}). For a low viewing angle with $\sin^2\gamma_{\rm view} \ll 1$ and $\langle a_{\rm max}\rangle \sim 0.25\,\rm \mu m$ constrained from the observed dust extinction in this study (Section \ref{sec:int_amax}), grains need to have high elongation $s \gg 1.4$ in order to reproduce the polarization efficiency at the maximum point.

Secondly, the size distribution of PAHs + Astrodust grains in the G11 filament has been constrained through the observed mean optical $\langle R_{\rm V}\rangle \sim 2.83$ from the optical dust extinction data by \cite{Zhang2025} - lower than the average $R_{\rm V} = 3.1$ in the ISM (\citealt{Cardelli1989}). The population of PAHs is considered to have the same fraction as that of Astrodust grains, so the variation $R_{\rm V}$ attribute to the change in Astrodust size distribution (i.e., depending on $a_{\rm max}$, see Figure \ref{fig:Rv}). Such a low $R_{\rm V}$ could be explained by the growth of PAHs (i.e., increasing the total mass fraction of PAHs $q_{\rm PAH}$) in the intermediate-density, translucent ISM, causing an increase in the extinction bump at 2175 $\AA$ and a lower $R_{\rm V}$ (\citealt{Zhang2025_pah}). The total PAH mass fraction $q_{\rm PAH}$ can be determined from its mid-IR emission and absorption features using modern spectroscopic instruments as the James Webb Telescope/Mid-IR Instrument (JWST/MIRI; see \citealt{Garci2022}; \citealt{Sturm2024}). Additionally, the extinction of magnetically aligned grains could vary with varying inclination angles between B-fields and the LOS, causing the variation of $R_{\rm V}$ with $\gamma_{\rm obs}$ (\citealt{Hensley2025}). We will soon update these effects of PAH growth and B-field inclination angles onto the dust extinction modeling in the upcoming version of \textsc{DustPOL\_py} code to improve further the constraints on intrinsic dust properties and 3D B-fields in local environments.


Despite the above uncertainties, the general bending trend of the B-fields toward the densest region of G11 inferred from the starlight polarization efficiency and alignment physics remains unaffected. This new technique unlocks a potential to study 3D B-fields and dust properties in the multiple scale of star-forming regions from current and future polarimetric observations. The new technique can be applied to obtain multi-scale 3D B-fields and dust properties using multi-wavelength dust polarization observations of nearby filamentary clouds such as the Polaris Flare (\citealt{Panopoulou2015}), L1544 (\citealt{Clemens2016}), and IC5146 (\citealt{Wang2017, Wang2020}), and distant filaments such as M17 SWex (\citealt{Sugitani2019}), G34.43+0.24 (\citealt{Soam2019}) and DR21 (\citealt{Ching2022}), to fully understand their variations and impacts during low-mass and high-mass filament formation and evolution. The investigation of 3D B-fields and dust properties in other filamentary clouds will be presented further in future studies of our 3D-BLISS series.

\section{Summary}
\label{sec:summary}
This paper first applies the new technique of inferring 3D inclination angles using observed $K_s$-band polarization of background stars by \cite{Chen2023} combined with RAT alignment theory (\citealt{TruongHoang2025}) to investigate 3D B-fields and dust properties in the G11 filament. The key findings are summarized as follows.
\begin{enumerate}
    \item We determined the magnetic alignment properties produced by RATs (e.g., the minimum aligned size $a_{\rm align}$ and the alignment function $f_{\rm align}(a)$) using the latest version of the \textsc{DustPOL\_py} code (\citealt{Lee2020}; \citealt{Tram2021, Tram2024}), giving the environmental conditions of G11 provided by multi-wavelength Herschel observations (\citealt{Zucker2018_filament}). 

    \item We used the interstellar dust extinction curve observed by the Gaia mission (\citealt{Zhang2025}) and the maximum polarization efficiency achieved in starlight polarization data at $K_s$ band by \cite{Chen2023} to provide constraints on intrinsic dust properties. Combining with the modeled extinction and polarization numerically predicted by the \textsc{DustPOL\_py} code, we found the mean maximum grain size of $\langle a_{\rm max} \rangle = 0.25\,\rm\mu m$ and the lower limit of grain elongation of $s \gtrsim 1.4$ presented in the outermost region of G11 with $N_{\rm H_2} \lesssim 10^{22}\,\rm cm^{-2}$. These intrinsic dust properties are assumed to be unchanged across the filamentary cloud.


    \item We determined the magnetic fluctuation effect, described by the turbulence factor $F_{\rm turb}$, from the observed POS polarization angle dispersion. The magnetic turbulence effect is minimal in the outer regions with $F_{\rm turb} \sim 0.8 - 0.9$ and becomes significant in high-density parts of G11 with lower $F_{\rm turb} < 0.5$.

    \item Once the effects of grain alignment, intrinsic dust properties, and magnetic turbulence are determined, the 3D inclination angles can be inferred from the observed starlight polarization efficiency $P_{\rm K}/N_{\rm H}$. The mean inclination angle is roughly $|\gamma_{\rm obs}| \sim 48^{\circ}$, but the local ones are highly scattered under the effect of 3D B-field variations across G11.

    \item The determination of 3D inclination angles is important for the estimation of 3D B-field strength and its roles in star formation processes. By taking the B-field inclination angles into account, the 3D B-field strength observed in G11 is higher than the POS one traditionally estimated from the DCF method. The dynamical importance of B-fields in regulating star formation is strengthened, resulting in lower \alfvenic Mach number $M_A$ and lower mass-to-flux ratio $\mu_{\phi}$.

    \item From the 2D observed POS polarization patterns and the inclination angles between the mean fields and the LOS, we reconstructed the variation of 3D B-field morphology within G11. We found the increase in inclination angles to the filament spine in Regions A and B, suggesting local arc-shaped B-fields with varying curvature toward the LOS. In Region C, we found a drop in inferred angles toward denser regions, which could be caused by the back-and-forth B-field bending by gravitational contraction. The origins of 3D B-field morphology and their roles in the formation and evolution of G11 can be further investigated by combining the 3D B-field observations from dust polarization with gas kinematics observations.

    \item Grain growth and PAH abundance as well as the B-field's inclination angle in the region of the maximum observed polarization could affect the accurate inference of B-field's inclination angle, but the inferred general bending of B-fields toward the densest region of G11 remains unaffected. Further analysis by combining available polarization with NIR/MIR dust extinction from Spitzer and JWST or spectral index from FIR/sub-mm thermal dust emission is required to accurately constrain dust properties and probe 3D B-fields on multiple scales of G11 and other IRDC filaments in the follow-up studies.

\end{enumerate}

\section*{Acknowledgements}
We thank the anonymous referee for helpful comments that improved the quality and the presentation of this paper. T.H. and B.T. acknowledge the support from the main research project (No. 2025186902) from Korea Astronomy and Space Science Institute (KASI). T.H. and N.B.N. acknowledge the support from the Vietnam National Foundation for Science and Technology Development (NAFOSTED) under grant numbers 
103.01-2025.147 and 103.99-2024.36. This research is based on archival $K_s$ band polarimetric observations by SIRPOL/ISRF at the South African Astronomical Observatory (SAAO), analyzed by \cite{Chen2023} and publicly available on the Science Data Bank (\citealt{Chen_data}). This paper has made use of archival data of interstellar dust extinction analyzed by \cite{Zhang2025}, obtained from optical extinction data from the European Space Agency (ESA) mission Gaia. This work has made use of NASA’s Astrophysics Data System. This work was partly supported by a grant from the Simons Foundation to IFIRSE, ICISE (916424, N.H.). We thank the ICISE staff for excellent support and hospitality.

\textit{Software:} Numpy (\citealt{Harris2020}), Scipy (\citealt{Virtanen2020}), Astropy (\citealt{Astropy2018}), DustPOL-py (\citealt{Lee2020}; \citealt{Tram2021, Tram2024}), ChatGPT(\citealt{ChatGPT}). We use the Matplotlib package (\citealt{Hunter2007}) for data visualization.

\textit{Facilities:} Simultaneous 3-color InfraRed Imager for Unbiased Survey (SIRUS) camera with the POLarimetry mode instrument (SIRPOL), Gaia Mission, Herschel Space Observatory.

\bibliography{ms}{}
\bibliographystyle{aasjournalv7}



\appendix

\section{Size distributions for Astrodust and Polycyclic Aromatic Hydrocarbons (PAHs)}
\label{sec:appendix_GSD}
In this section, we discuss in detail our assumptions of the size distributions for Astrodust and PAHs used in this study. For Astrodust grains, it follows the power-law size distribution (\citealt{Mathis1977}), with a power-law index of $\beta$ as
\begin{equation}
    n_{\rm Astro}(a)da = C_{\rm Astro}a^{-\beta}da, 
\end{equation}
where $C_{\rm Astro}$ is the normalization constant of Astrodust grains derived from the dust-to-gas mass ratio $M_{d/g}$ as (see \citealt{Tram2020})
\begin{equation}
    C_{\rm Astro} = \frac{(4 + \beta)M_{d/g}m_{\rm gas}}{\frac{4}{3}\pi\rho_{\rm Astro}(a_{\rm max}^{4 + \beta} - a_{\rm min}^{4 + \beta})},
\end{equation}
where $m_{\rm gas} = 1\,\rm amu$ is the hydrogen mass, and $\rho_{\rm Astro} = 2.74\,\g\cm^{-3}$ is the mass density of Astrodust grains with a porosity of 0.2 (\citealt{Hensley2023}). We consider a typical power-law index of $\beta = 3.5$ and dust-to-gas mass ratio of 0.01 in the diffuse ISM (\citealt{Mathis1977}). The assumption of power-law distribution is reasonable for large grains found in the dense filaments and starless cores  (\citealt{Hirashita2012}; \citealt{Bate2022}).

For PAH size distribution, we consider it follows the two log-normal distributions described by \cite{DraineLi2007} as
\begin{equation}
    n_{\rm PAH}(a)da = \sum_{j = 1}^2 \frac{n_{0,j}}{a} \exp\left(-\frac{[\ln(a/a_{0,j}^{\rm PAH})]^2}{2 \sigma^2_{\rm PAH, j}}\right) da, 
\end{equation}
where $a^{\rm PAH}_{0,j}$ and $\sigma_{\rm PAH, j}$ are the peaks and widths in each log-normal component. Above that, the factor $n_{0,j}$ is determined from the total abundance of carbon atoms per H nucleus in each log-normal component, $b_j$, as
\begin{equation}
    n_{0,j} = \frac{3}{(2\pi)^{3/2}}\frac{\exp(4.5\sigma^2_{\rm PAH, j})}{1 + \textrm{erf}(x_j)}\frac{m_C}{\rho_C a^3_{M, j}\sigma_{\rm PAH, j}}b_j,
\end{equation}
with
\begin{equation}
    x_j = \frac{\ln(a_{M, j}/a_{\rm min})}{\sqrt{2}\sigma_{\rm PAH, j}},
\end{equation}
where $\rho_C = 2.26\,\rm g\,cm^{-3}$ is the carbon mass density, $m_C = 12.011\,\rm u$ is the carbon mass, and $a_{M, j} \equiv a_{0,j}^{\rm PAH}\exp{(3\sigma^2_{\rm PAH, j})}$ is the location of the peak in the mass distribution ($\varpropto a^3 dn/d\ln a$). We adopt $a^{\rm PAH}_{0,j}$, $\sigma_{\rm PAH, j}$ and $b_j$ of each log-normal component constrained from mid-IR emission of interstellar dust (\citealt{DraineLi2007}). The properties of the size distribution for Astrodust and PAHs are summarized in Table \ref{tab:model_paramater}.

\begin{table*}
    \centering
    \caption{List of modeled parameters for Astrodust + PAHs size distribution}
    \label{tab:model_paramater}
    \begin{tabular}{c c c c}
         \toprule
          Parameters & Values & Meaning & References\\
         \midrule
          $a_{\rm min}$ & $3.5\,\rm \AA$ & Minimum grain size & (1), (2)\\
          $M_{d/g}$ & 0.01 & Dust-to-gas mass ratio & (3)\\
          $\beta$ & 3.5 & Power-law index of the size distribution for Astrodust & (3)\\
          $b_{1}$ & 45 ppm & Carbon abundance per H nuclei of the log-normal component 1 of PAH size distribution & (2), (4)\\
          $b_{2}$ & 15 ppm & Carbon abundance per H nuclei of the log-normal component 2 of PAH size distribution & (2), (4)\\
          $a_{0,1}^{\rm PAH}$ & $4\,\rm \AA$ & Peak of the log-normal component 1 of PAH size distribution & (2), (4)\\
          $a_{0,2}^{\rm PAH}$ & $20\,\rm \AA$ & Peak of the log-normal component 2 of PAH size distribution & (2), (4)\\
          $\sigma_{\rm PAH, 1}$ & 0.4 & Width of the log-normal component 1 of PAH size distribution & (2), (4)\\
          $\sigma_{\rm PAH, 1}$ & 0.55 & Width of the log-normal component 2 of PAH size distribution & (2), (4)\\
       
         \bottomrule
    \end{tabular}

      \textbf{References:} (1) \cite{Draine1979}; (2) \cite{DraineLi2007}; (3) \cite{Mathis1977}; (4) \cite{Wang2015}
\end{table*}

\section{Magnetic alignment of dust grains by MRAT}
\label{sec:appendix_RAT}
In this section, we present the properties of grain alignment with respect to B-fields induced by MRAT (\citealt{Hoang2016a}). Dust grains within the filamentary clouds can be spun up to suprathermal rotation and be aligned with B-fields by RATs from ambient radiation fields (see \citealt{Lazarian2007}; \citealt{Hoang2014}; \citealt{Hoang2016a}). The minimum size in which grains can achieve magnetic alignment by RATs, $a_{\rm align}$, is calculated by (see \citealt{Hoang2021})
\begin{equation}
\label{eq:align_ana}
\begin{split}
     a_{\rm align} &\simeq 0.055\hat{\rho}^{-1/7} \left(\frac{\gamma_{\rm rad}U}{0.1}\right)^{-2/7}\left(\frac{n_\H}{10^3\,\cm^{-3}}\right)^{2/7}\left(\frac{T_{\rm gas}}{10\,\rm K}\right)^{2/7} \left(\frac{\bar{\lambda}}{1.2\mum}\right)^{4/7} (1+F_{\rm IR})^{2/7} ~\mum,  
\end{split}
\end{equation} 
where $\hat{\rho}=\rho/(3\g\cm^{-3})$ is the dust mass density, $T_{\rm gas}$ is the gas temperature, $n_{\H}$ is the number density of hydrogen atoms, $\gamma_{\rm rad}$ is the anisotropy degree of radiation, $\bar{\lambda}$ is the mean wavelength of the radiation field, $U$ is the radiation field strength with respect to the interstellar radiation field (ISRF) in the solar neighborhood \citep{Mathis1983}, and $F_{\rm IR}$ is a dimensionless parameter that describes the grain rotational damping by IR emission. In the cold, dense environments as filaments and dense cores, the damping by IR emission is minimal, giving $F_{\rm IR} \ll 1$. Once the radiation (i.e., $U$ and $\gamma_{\rm rad}$) and gas properties (i.e., $n_{\rm H}$ and $T_{\rm gas}$) are provided, the minimum aligned size $a_{\rm align}$ in local environments can be calculated as shown in Equation \ref{eq:align_ana} (see Figure \ref{fig:align}). Then, we can calculate the alignment function $f_{\rm align}(a)$ as shown in Equation \ref{eq:falign}, considering $R = 1$ for large grains $a > a_{\rm align}$ in this study (see Section \ref{sec:align}).

From the modern alignment theory based on RATs, the Rayleigh reduction factor $R$ is governed by the alignment properties of grains at high-J (i.e., rotating suprathermally by RATs with $\Omega \gg \Omega_{\rm ther}$) and low-J attractors (i.e., rotating subthermally by gas collision with  $\Omega \sim \Omega_{\rm ther}$), which is given by (see \citealt{Hoang2016a})
\bea
R=f_{\rm high-J}Q_{X}^{\rm high-J}+ (1-f_{\rm high-J})Q_{X}^{\rm low-J},\label{eq:Ralign}
\ena
where $Q_{X}^{\rm high-J}$ and $Q_{X}^{\rm low-J}$ are the degree of alignment for grains at high-J and low-J attractors, respectively. Numerical calculations in \cite{Hoang2016a} showed that grains at high-J attractors can be perfectly aligned with ambient B-fields by RATs with $Q_{X}^{\rm high-J} = 1$, while grains at low-J attractors have smaller alignment degree with $Q_{X}^{\rm low-J} \sim 0.3 $ \citep{Hoang2016a}.

Above, $f_{\rm high-J}$ is the fraction of grains at high-J attractors. For paramagnetic (PM) grains, a small portion of grains are at high-J attractor points for a limited range of the radiation direction (\citealt{Hoang2016a}). For superparamagnetic (SPM) grains with iron inclusions, $f_{\rm high-J}$ is found to increase with the strength of magnetic relaxation over the gas randomization, denoted by $\delta_{\rm mag}$ (\citealt{Hoang2016a}). The fraction $f_{\rm high-J}$ for SPM grains is parameterized as a function of $\delta_{\rm mag}$ as (see \citealt{Giang2023})
\bea 
f_{\rm high-J}(\delta_{\rm mag}) = \left\{
\begin{array}{l l}    
    0.25 ~ ~  {\rm ~ for~ } \delta_{\rm mag} < 1   \\
    0.5 ~ ~  {\rm ~ for~ } 1 \leq \delta_{\rm mag} \leq 10 \\
    1    ~ ~ ~ ~  {\rm ~ for~}  \delta_{\rm mag} > 10 \\
\end{array}\right. ,
\label{eq:fhiJ_deltam}
\ena
with 
\bea
\delta_{\rm mag} = 5.6a_{-5}^{-1}\frac{N_{\rm cl}\phi_{\rm sp,-2}\hat{p}^{2}B_{2}^{2}}{\hat{\rho} n_{3}T_{\gas,1}^{1/2}}\frac{k_{\rm sp}(\Omega)}{T_{\rm d,1}},\label{eq:delta_m}~~~~
\ena
where $B_{2}=B/10^{2}\mu G$ is the normalized magnetic field strength, $n_{3}=n_{\rm H}/10^3\cm^{-3}$ is the normalized gas density, $T_{\rm g,1}=T_{\rm gas}/10\K$ and $T_{\rm d,1}=T_{\rm d}/10\K$ is the normalized gas and dust temperature, respectively. $N_{\rm cl}$ and $\phi_{\rm sp,-2} = \phi_{\rm sp}/10^{-2}$ are the number of iron inclusions and the volume filling factor, $a_{-5} = a /10^{-5}\,\rm cm$ and $\hat{p} = p/5.5$. $k_{\rm sp}(\Omega)$ is the function of the grain angular velocity $\Omega$ describing the suppression of the magnetic susceptibility at high angular velocity (see \citealt{Hoang2022_Graingrowth}). The presence of iron inclusions inside SPM grains can enhance the grain's magnetic susceptibility and the magnetic relaxation strength; thus, increases the RAT alignment efficiency (see \citealt{Lazarian2008}; \citealt{Hoang2016a}). Once the magnetic properties of SPM grains (i.e., $N_{\rm cl}$ and $\phi_{\rm sp}$) and the properties of local environments (i.e., $B$, $n_{\rm H}$, $T_{\rm gas}$ and $T_d$) are identified, we can calculate $\delta_{\rm mag}$, $f_{\rm high-J}$ and the exact value of $R$ for large SPM grains $a > a_{\rm align}$ with MRAT alignment.

\section{Effect of magnetic turbulence on the starlight depolarization}
\label{sec:appendix_Fturb}
The perpendicular fluctuations of local B-fields $\bf \delta B_{\perp}$ with respect to the mean fields $\bf B_0$ (i.e., \alfvenic turbulence) could contribute to the depolarization presented in the observations of the diffuse ISM and dense filamentary clouds (see, e.g., \citealt{Lee1985}; \citealt{Planck2020}; \citealt{Tram2024}). \cite{TruongHoang2025} demonstrated that the effect of magnetic fluctuation on the observed starlight polarization degree can be characterized by the turbulence factor $F_{\rm turb}$, which is calculated as (see also \citealt{Lee1985}) 

\begin{equation}
    F_{\rm turb} = 1 - 1.5\sin^2(\Delta\theta),
\label{eq:Fturb}
\end{equation}
where $\Delta\theta$ is the deviation angle between the local and mean B-fields in the 3D space. For sub-\alfvenic turbulence and $\Delta\theta < 20.5$ degrees, $F_{\rm turb} \approx 1 - 1.5(\Delta\theta)^2$ (see the detailed analysis in \citealt{TruongHoang2025}). Lower $F_{\rm turb}$ corresponds to the strong magnetic turbulence contributing to the decrease in starlight polarization efficiency.

\section{Degeneracy between the intrinsic dust properties and 3D inclination angles inferred from dust polarization}
\label{sec:appendix_degeneracy}
In this section, we demonstrate quantitatively how the intrinsic dust properties can modify the inference of 3D inclination angles from current observed starlight polarization efficiency $P_{\rm K}/N_H$ by \cite{Chen2023}. We vary the values of maximum grain size $a_{\rm max}$ and grain elongation $s$ in the outer regions of G11. Then, we re-calculate the intrinsic polarization efficiency $P_{\rm i,K}/N_H$ (Equation \ref{eq:Pi}) and the polarization coefficient fraction $f_{\rm pol,K}$ (Equation \ref{eq:fpol}), and re-determine the inferred inclination angles (Equation \ref{eq:chi_ext}).

\begin{figure*}
    \includegraphics[width=0.5\linewidth]{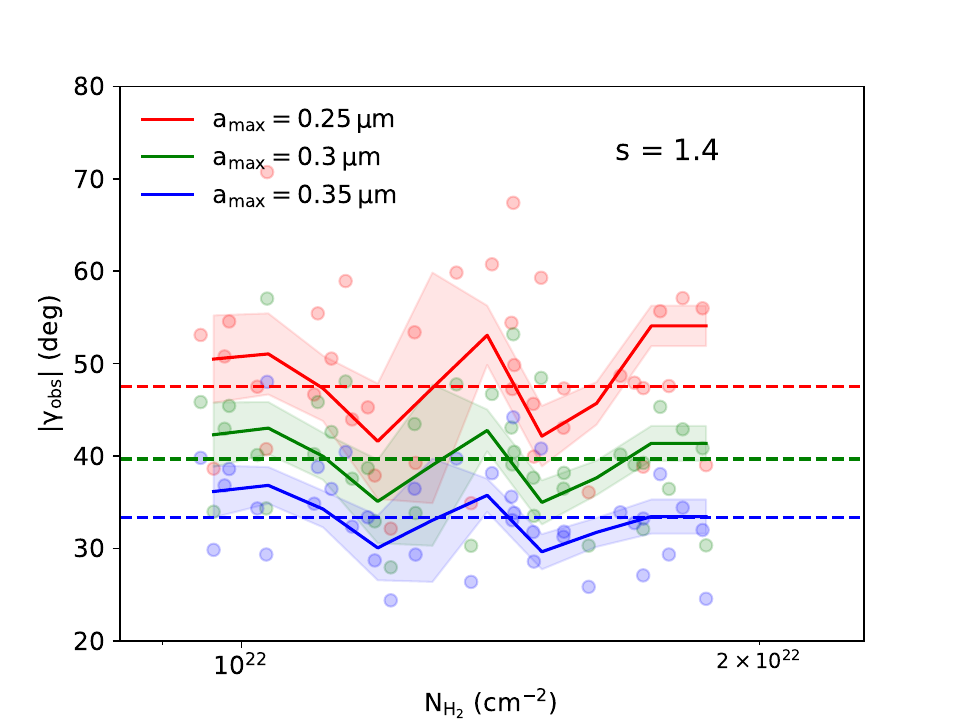}
    \includegraphics[width=0.5\linewidth]{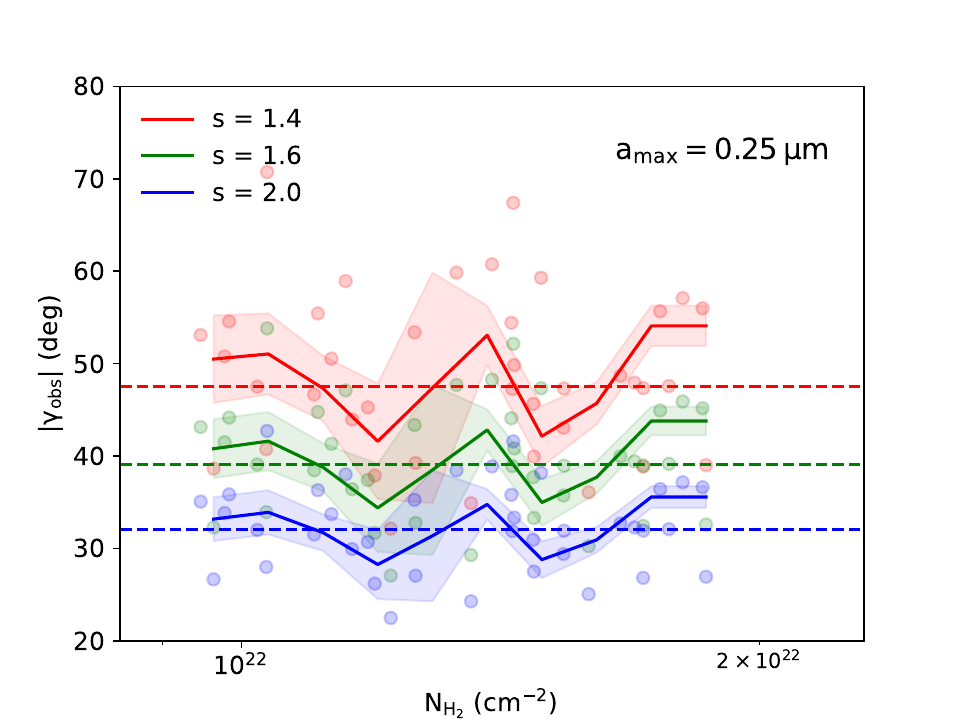}
    \caption{The profile $|\gamma_{\rm obs}| - N_{\rm H_2}$ similar to the right panel of Figure \ref{fig:Incl_Star}, but considering increasing $a_{\rm max}$ (left panel) or increasing grain elongation $s$ (right panel).}
    \label{fig:Incl_degeneracy}
\end{figure*}

Figure \ref{fig:Incl_degeneracy} shows the dependence of the profile $|\gamma_{\rm obs}| - N_{\rm H_2}$ on the increasing $a_{\rm max} $ (left panel) and the increasing grain axial ratio $s$ (right panel). Once the maximum grain size $a_{\rm max}$ increases to $0.35\,\rm\mu m$, the values of inferred angles $|\gamma_{\rm obs}|$ decrease by a factor of $\sim 1.6$. This effect occurs similarly when the grain elongation increases from $s = 1.4$ to $s = 2$ (i.e., more elongated). The variation of $|\gamma_{\rm obs}|$ with increasing $N_{\rm H_2}$ does not change significantly since the intrinsic dust properties are considered to be constant across G11 throughout this study. Our conclusion about the inferred general trend of B-field bending across the cloud remains unchanged (see Figures \ref{fig:incl_filament} and \ref{fig:3D_G11}).

\section{The dependence of extinction curve $A_{\lambda}/N_{\rm H}$ on intrinsic dust properties}
\label{sec:appendix_extinction}

In this section, we present the variation of optical-NIR dust extinction curve with the intrinsic dust properties (i.e., $a_{\rm max}$ and grain axial ratios $s$). The extinction magnitude normalized to the hydrogen column density, denoted by $A_{\lambda}/N_{\rm H}$, is given by
\begin{equation}
    \frac{A_{\lambda}}{N_{\rm H}} = \sum_{\rm j = Astro, PAH}\int_{a_{\rm min}}^{a_{\rm max}}C_{\rm ext, j}(a,\lambda) n_{j}(a)da,
\label{eq:Alambda}
\end{equation}
where $n_{j}(a)$ is the size distribution of the dust component $j$. Above, $C_{\rm ext, j}(a, \lambda)$ is the total extinction cross-section of the grain size $a$ is each dust component $j$ and is calculated by 
\begin{equation}
    C_{\rm ext,j}(a, \lambda) = \frac{1}{3}\,[2C_{\rm ext, j}(a, \lambda)(\bE \perp \ba_{1}) + C_{\rm ext, j}(a, \lambda)(\bE \parallel \ba_{1})],
\end{equation}
where $C_{\rm ext,j}(a, \lambda)(\bE \perp \ba_{1})$ and $C_{\rm ext,j}(a, \lambda)(\bE \parallel \ba_{1})$ are obtained from Astrodust grains with different elongations (\citealt{Draine2021}) and PAHs found in the diffuse ISM (\citealt{Hensley2023}).

\begin{figure*}
    \includegraphics[width=1.0\linewidth]{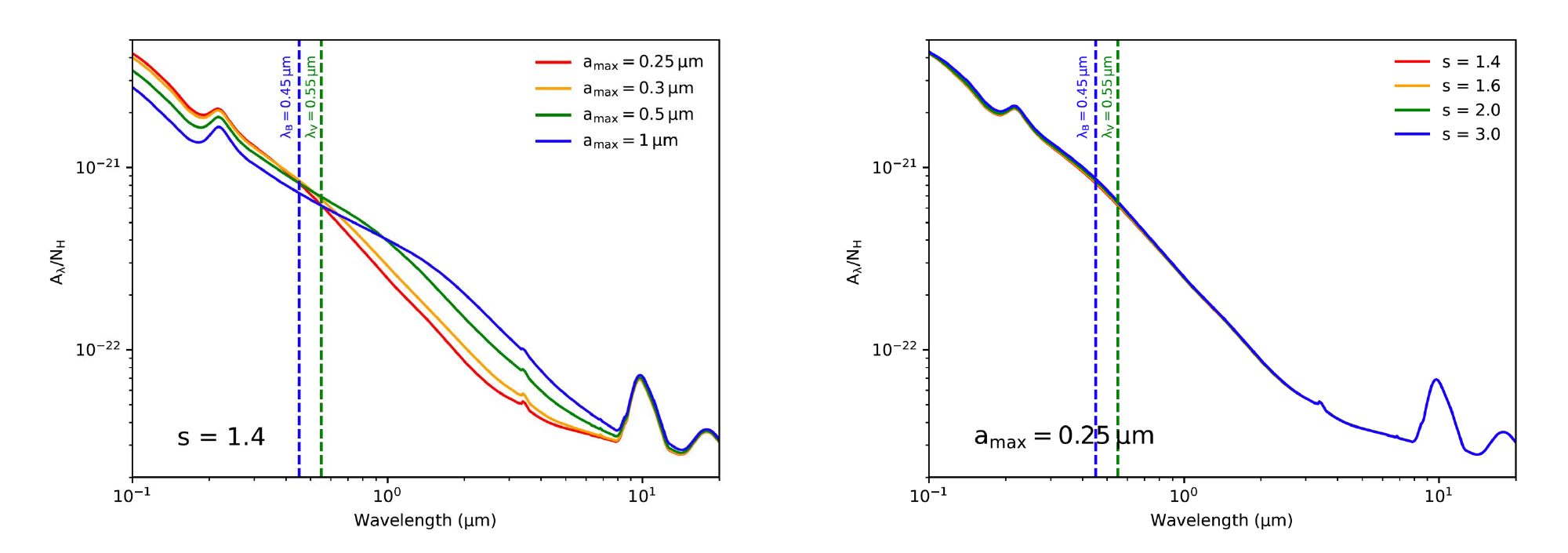}
    \caption{The normalized dust extinction curve $A_{\lambda}/N_{\rm H}$ at optical-NIR wavelengths ($\lambda = 0.1 - 20\,\rm\mu m$) modeled by \textsc{DustPOL\_py} code for Astrodust+PAHs grains. Varying $a_{\rm max}$ (left panel) and axial ratios of oblate grains (right panel) are assumed.}
    \label{fig:extinction}
\end{figure*}

We perform the numerical calculation of optical-NIR dust extinction $A_{\lambda}/N_{\rm H}$ as shown in Equation \ref{eq:Alambda}, using the latest version of the \textsc{DustPOL\_py} code (see \citealt{Tram2025}). Figure \ref{fig:extinction} shows the normalized extinction curve $A_{\lambda}/N_{\rm H}$ produced by Astrodust+PAHs with respect to the observed wavelength from optical to NIR ($\lambda = 0.1 - 20\,\rm\mu m$), considering varying $a_{\rm max} = 0.25 - 1.0\,\rm\mu m$ (left panel) and grain elongation $s = 1.4 - 3$ (right panel). For larger $a_{\rm max}$, the enhanced population of larger grain sizes causes the lower extinction at shorter wavelengths and the higher extinction at longer wavelengths, leading to a flatter extinction curve and higher $R_{\rm V}$ derived at B- and V-bands (see Section \ref{sec:int_amax}). The grain elongation has less impact on the dust extinction, and thus, the shape of the extinction curve is unchanged for different grain axial ratios. 

\end{document}